%% file: main.tex
\documentclass[runningheads]{article}
\usepackage[T1]{fontenc}
\usepackage{authblk}
\usepackage{orcidlink}

\usepackage{colortbl}
\usepackage{xspace}
\usepackage{array}
\usepackage{tabularx}

\usepackage{graphicx}
\usepackage{amsmath}
\usepackage{hyperref}

\usepackage{subcaption}

\makeatletter
\renewcommand\p@subfigure{} 
\makeatother
\usepackage[all=normal,wordspacing=normal,mathspacing=tight,tracking=tight,paragraphs=tight,floats=tight]{savetrees}

\usepackage{booktabs}
\usepackage[table,x11names,dvipsnames,table]{xcolor}

\definecolor{lightgray}{gray}{0.9}

\usepackage{cite}

\usepackage{url}

\begin{document}

\title{Routing Attacks in Ethereum PoS:\\ A Systematic Exploration}

\author{Constantine Doumanidis\orcidlink{0000-0003-2479-8187}}
\author{Maria Apostolaki\orcidlink{0000-0003-0342-2631}}
\affil{Princeton University}

\maketitle

\newcommand{\myitem}[1]{\vspace*{0.035in}\noindent\textbf{#1}}
\newcommand{\eg}{{\em e.g., }}
\newcommand{\ie}{{\em i.e., }}

\newcommand{\attackone}{\textsc{StakeBleed}\xspace}
\newcommand{\attacktwo}{\textsc{KnockBlock}\xspace}
\definecolor{darkgreen}{rgb}{0.13, 0.55, 0.13}

\newcommand{\node}{\textsc{SuperNode}\xspace}

\input{sections/00_abstract}
\input{sections/01_introduction}
\input{sections/02-back}
\input{sections/03i_overview}
\input{sections/04_identifying_stakeholders}
\input{sections/05_topoanalysis}
\input{sections/05_attackone}
\input{sections/05_attacktwo}
\input{sections/09_countermeasures}
\input{sections/11_related_work}
\input{sections/12_conclusion}
\input{sections/15_acks}

\bibliographystyle{IEEEtran}
\bibliography{bibliography}

\input{sections/14_appendix}

\end{document}

%% file: sections/00_abstract.tex
\begin{abstract}

With the promise of greater decentralization and sustainability, Ethereum transitioned from a Proof-of-Work (PoW) to a Proof-of-Stake (PoS) consensus mechanism. 
The goal of this paper is to investigate the security of Ethereum’s PoS system from an Internet routing perspective.
To this end, this paper makes two contributions:
First, we devise a novel framework for inferring the distribution of validators on the Internet with minimal disturbance to the real network.
Second, we introduce a class of network-level attacks on Ethereum’s PoS system that jointly exploit Internet routing vulnerabilities with the protocol’s reward and penalty mechanisms. We describe two representative attacks: \attackone, where the attacker triggers an inactivity leak, halting block finality and causing financial losses for all validators; and \attacktwo, where the attacker increases her expected MEV gains by preventing targeted blocks from being included in the chain.
We find that both attacks are practical and effective. An attacker executing \attackone can inflict losses of almost 300 ETH in just 2 hours by hijacking as few as 30 IP prefixes.
An attacker implementing \attacktwo could increase their MEV expected gains by 44.5\%  while hijacking a single prefix for less than 2 minutes.
Finally, we discuss countermeasures practical for validator operators, consensus and P2P protocol designers.
Our paper serves as a call to action for validators to reinforce their Internet routing infrastructure and for the Ethereum P2P protocol to implement stronger mechanisms to conceal validator locations.

\end{abstract}

%% file: sections/01_introduction.tex
\section{Introduction}
Ethereum and many other blockchain systems rely on Proof-of-Stake (PoS) consensus to maintain a globally consistent ledger. Validators propose and attest to blocks via a peer-to-peer (P2P) network, but because this communication traverses the public Internet, it remains exposed to routing attacks such as BGP hijacking which allows adversaries to redirect Internet traffic through their own infrastructure.

\myitem{Practicality:}
BGP hijacking remains a practical threat; despite the growing adoption of RPKI~\cite{cryptoeprint:2016/1010}, thousands of hijacks~\cite{qrator2023q1,qrator2023q2,qrator2025q1,manrs2022bgp} are observed each month that even affect IP prefixes protected by RPKI~\cite{cloudflare2024incident}. 
Critically, as we discuss in this paper, 40.85\% of Ethereum nodes are not (properly) protected by RPKI. 
Beyond the lack of a comprehensive prevention mechanism, two factors make BGP hijacks a particularly practical attack. First, resolving hijacks after they have been detected is a slow manual process, taking at least 2-3 hours even for expert teams~\cite{cloudflare2024incident,cloudflare2021facebook}. Second, legal repercussions are rare for the attacker, as hijacks are indistinguishable from misconfigurations, which are common~\cite{mahajan2002understanding, cloudflare2021facebook}.
Recent incidents targeting the Celer Bridge~\cite{celer} and MyEtherWallet~\cite{myetherwallet} further underscore the practicality of BGP hijacking in general and against blockchains. These attacks lasted hours, resulted in multi-million dollar gains despite RPKI, and incurred no consequences for the attackers.

Investigating the effect of BGP hijacking on Ethereum PoS requires contributions in two key directions: \emph{(i)} understanding how PoS consensus properties such as liveness and safety can be threatened by such network vulnerabilities; and \emph{(ii)} inferring Ethereum’s network-level topology to assess how key contributors of the protocol are distributed and exposed to such attacks.

\myitem{Attacking Ethereum PoS with BGP hijacks:}
We introduce two representative attacks against Ethereum’s Proof-of-Stake (PoS) protocol: \attackone and \attacktwo. In both cases, the attacker controls a BGP-speaking router on the Internet—either through a malicious Internet Service Provider (ISP) or by compromising the infrastructure of a legitimate one (as in ~\cite{orange2024bgp, wired2014bgphijack}) or by setting up their own Autonomous System (AS)~\cite{own-asn}. The two attacks differ in their target selection, level of stealth, and potential impact.

In \attackone a network adversary seeks to isolate a set of validators from the rest of the network, preventing them from contributing to consensus. In the worst case, \attackone can stop block finality, leaving the chain in an uncertain state where transactions may be reverted and critical applications stall. Even when the attack is not fully successful, it incurs financial losses at both isolated and non-isolated validators. 

In \attacktwo the network attacker aims to prevent a specific block proposer from broadcasting its block. If the attacker controls the next block proposer, she directly benefits from the attack, as the missed block creates additional MEV opportunities and increases the attestation rewards for that block. This attack is highly stealthy, requiring the hijack to be effective for only a few seconds.\footnote{Approximately 2 minutes when also accounting for BGP advertisement propagation time.}

\myitem{Novelty:} While prior work has leveraged network-layer vulnerabilities—particularly BGP hijacks~\cite{HKZG_15,heo2023partitioning,tran2020stealthier,saad2021syncattack}—their impact has largely been confined to PoW-based systems. To our knowledge, this is the first work to exploit Ethereum’s PoS-specific characteristics, enabling both direct financial gain and inflicting financial damage \textemdash whereas prior PoW attacks focused primarily on reputational harm~\cite{apostolaki2017hijacking, tran2024routing, saad2021syncattack}. 
By contrast, recent attacks on PoS Ethereum~\cite{neuder2021low, neu2021ebb, three-attacks-on-pos, two-more-attacks, pos-under-scrutiny} do not exploit Internet routing vulnerabilities.

\myitem{Inferring network-level topology of validators:} We design a multi-stage algorithm to infer the network-layer locations of Ethereum validators, enabling a realistic assessment of Ethereum’s susceptibility to network-layer attacks and helping operators understand what behaviors may inadvertently reveal their validators.
Our algorithm combines a scalable collector (\node) with a multi-stage inference pipeline:
(i) \node passively accepts inbound connections, maintaining them just long enough to determine whether the peer hosts a validator. It \emph{selectively} logs the timing of various types of attestations \textemdash along with auxiliary signals such as eagerly pushed or unsolicited messages\textemdash to reliably infer validator identities while minimizing resource overhead.
(ii) High-confidence mappings are bootstrapped from validators that are activated on the blockchain in sequence and co-located, reflecting a common deployment pattern by a single operator.
(iii) These mappings seed a machine learning model that generalizes timing patterns to the broader validator set, allowing us to map it in its entirety. 
This is possible because, while sequential activation and co-location of validators is not a uniform pattern (\eg not all validators are activated at the same time, and even those that are might be distributed across locations), the propagation pattern of these nodes is representative of that of any validator node as it reflects protocol-level behavior rather than operator-specific idiosyncrasies.
We validate our methodology by verifying that co-located validators are not common in non-consecutive validator IDs, and by observing that validators belonging to distinct Lido operators are often co-hosted with their operators' websites.

\myitem{Novelty:} 
Similar analyses exist for PoW-based Ethereum and Bitcoin, but their methodologies do not carry over to PoS. Prior research identified miners—similarly influential nodes—either through the websites of their public Stratum servers~\cite{tran2024routing} or by analyzing block propagation delays using simple heuristics~\cite{apostolaki2017hijacking}, neither of which translates to our setting.
The difficulty of constructing a comprehensive validator-to-node mapping in PoS Ethereum is further underscored by recent work, which was only able to infer $\approx15\%$ of validators~\cite{heimbach2024deanonymizing}. This work is the first to provide a full mapping.

\myitem{Countermeasures:} 
While securing the entire Internet is the ideal solution to such attacks, it is not within the capabilities of the Ethereum and broader blockchain ecosystem. Hence, we propose countermeasures at the deployment, consensus, and peer-to-peer layers, enabling validator operators, protocol designers, and P2P developers to respond within their respective domains. 

\myitem{Ethics:} No real Ethereum clients or IP prefixes were targeted in this study. We have disclosed our findings to the Ethereum Foundation. For details, refer to Appendix \ref{sec:ethics}.

%% file: sections/02-back.tex
\section{Preliminaries}
\label{sec:preliminaries}

\myitem{Ethereum’s Proof-of-Stake (PoS) system} requires validators to deposit 32 ETH and maintain high availability to perform time-sensitive duties that secure the network. In exchange, validators earn rewards proportional to their stake. Validators operate through a set of interconnected clients: the execution client handles transactions and smart contracts, the consensus client manages the PoS logic, and the validator client submits attestations and proposes blocks. Each node in the network is identified via an Ethereum Node Record (ENR), and node participation in the network is dynamic, with frequent churn—the process of nodes entering and exiting the network. At the protocol level, validators contribute to network security and liveness by issuing attestations (votes on block validity and network state) and occasionally proposing blocks that extend the blockchain. Validators are also assigned to committees, which determine which parts of the network they interact with and which messages they are expected to sign.

\myitem{libp2p} is a modular P2P networking stack used by Ethereum. It includes a pubsub system that allows nodes to subscribe to and publish messages in specific topics. Validators subscribe to topics based on their committee assignments, ensuring efficient and relevant message dissemination. Ethereum primarily uses Gossipsub~\cite{vyzovitis2020gossipsub}, a gossip-based pubsub protocol that maintains a mesh of connections per topic. Gossipsub distinguishes between long-lived (mesh) and short-lived (fanout) peers, and uses heartbeats to maintain mesh health. It relies on several control messages: GRAFT and PRUNE adjust peer connections, while IHAVE and IWANT help peers exchange metadata about unseen messages, ensuring reliable propagation without overwhelming the network.

\myitem{The Border Gateway Protocol (BGP)}~\cite{rekhter2006border} is the Internet’s de facto interdomain routing protocol, governing how IP packets are forwarded across Autonomous Systems. ASes exchange routes for IP prefixes, each of which is originated by a single AS and propagated through the network. Routers independently select next hops based on routes advertised by their neighbors. BGP does not verify route authenticity by default, allowing any AS to advertise false routing information. Such false advertisements, known as BGP hijacks, can misdirect traffic to unintended destinations and are a powerful mechanism for traffic interception attacks.

%% file: sections/03i_overview.tex
\section{Overview}

\subsection{Threat model}
\label{ssec:threat}
Our threat model follows prior routing-adversary work~\cite{tran2024routing, threebirds2023}. The attacker controls a BGP-speaking router—either by compromising~\cite{orange2024bgp, wired2014bgphijack}, acquiring, or setting up an AS~\cite{own-asn, nick-isp}. She performs BGP hijacks (more-specific/subprefix or same-length) targeting prefixes hosting Ethereum nodes. More-specific / forged-origin subprefix announcements succeed because routers prefer longest-prefix match; same-length hijacks bypass origin-based RPKI checks and can still divert a large fraction of traffic depending on location. We find 40.85\% of Ethereum nodes are in prefixes lacking RPKI protection: 26.98\% are vulnerable due to permissive maxLength settings and 13.87\% lack ROAs and have short prefixes. Even RPKI-protected prefixes remain practically vulnerable unless downstream ISPs consistently filter forged advertisements—a responsibility often unmet in practice~\cite{cloudflare2024incident}. Same-length hijacks also remain effective, typically diverting on the order of 50\% of a victim’s connections on average~\cite{sun2021securing,10.1145/2659899}.

To reduce visibility, the attacker limits hijack duration and scope, with specifics depending on the attack goal. Critically, though, unless the attacker stops the hijack (as in \cite{klayswap_bgp}), it will be effective for at least 2-3 hours, as resolving it is a human-driven process~\cite{cloudflare2021facebook, cloudflare2024incident, celerbridge_bgp, myetherwallet} by the network operator (ISP) rather than affected applications.

\begin{figure*}
    \centering
    \includegraphics[width=0.96\linewidth, trim={0cm 5.45cm 0 0},clip]{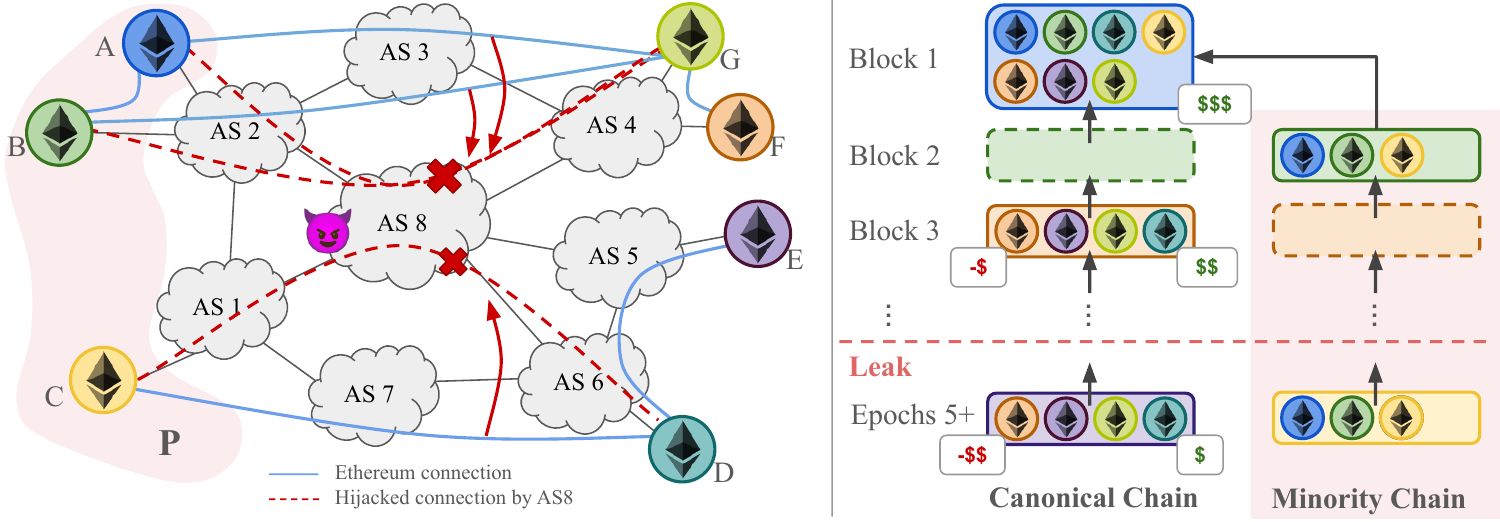}
    \caption{\textbf{Network view (left)}: A \attackone adversary controlling a router in AS 8 isolates nodes in $P$, from the rest of the Ethereum network by diverting their traffic using a BGP hijack and dropping their connections to nodes outside $P$.\\ \textbf{Blockchain view (right)}: Nodes in $P$ have a minority chain view of the blockchain.
    Before the inactivity leak is triggered, validators in $P$ lose proposer rewards and incur penalties because they do not produce attestations in the canonical chain, while validators outside $P$ also lose some of their rewards from proposing and attesting. After 4 epochs of $P$ being isolated, the inactivity leak is triggered, causing validators in $P$ to incur inactivity penalties, while validators outside $P$ stop receiving attestation rewards.}
    \label{fig:stakebleed-overview}
\end{figure*}

\subsection{\attackone}
\label{sec:stakebleed_selecting_p}
In \attackone, the adversary aims to isolate a set of validators $P$ from the rest of the network, preventing them from contributing to consensus. If $P$ controls more than one-third of the total stake, the attack halts block finality, leaving the chain in an uncertain state in which transactions may be reverted, critical applications stall, and all validators incur significant financial losses from missed rewards, and penalties.

\myitem{Description:} The attacker first maps validator keys to IP addresses, e.g., using the algorithm in \S\ref{sec:identifying_stakeholders}, and then launches BGP hijacks targeting the most specific prefixes of some of those IPs to divert traffic to a selected set of nodes $P$. 
To maintain stealth, she prioritizes prefixes with higher validator density, adding them to $P$ until the 33\% stake threshold is met. After initiating the hijack, the attacker monitors the chain for attestations from targeted validators; if some appear due to mapping errors or ineffective hijacks, she expands $P$ iteratively.

Once on-path, the attacker inspects Ethereum traffic (e.g., via known ports) and selectively drops packets to enforce the partition. Hijacks must persist for at least 4 epochs (128 slots / $\approx 26$ minutes) to trigger an inactivity leak, a mechanism in Ethereum PoS for handling absent stake without compromising consistency or safety. During a leak, blocks are not finalized, and attestation rewards cease. The attack tolerates imperfect mappings or partial failures, as $P$ can be dynamically adjusted. For a detailed walk-through of \attackone on Fig.~\ref{fig:stakebleed-overview}, see \S\ref{apdx:stakebleed_example}.

\myitem{Impact:}
If $|P| > 33\%$, block finalization stops, putting the blockchain in a state of uncertainty where transactions—even those included in blocks—are not guaranteed to be irreversible. This undermines user trust, halts critical applications (e.g., DeFi protocols that rely on finalized state), and may trigger cascading effects like stalled bridges, paused staking services, and halted validator exits.

Furthermore, validators both in $P$ and outside $P$ are harmed financially. 
During the attack, validators in $P$ miss all rewards, 
and during
a leak accrue inactivity scores, leading to inactivity penalties that grow over time. For $|P|=35\%$ and a two-hour attack, losses total 110.6 ETH in penalties and 61.8 ETH in missed rewards.
Validators outside $P$ are also harmed, as their attestations are delayed or omitted, and blocks include fewer attestations -- for $|P|=35\%$, this leads to an additional 114.8 ETH in aggregate losses.
Even without reaching the inactivity leak threshold, the attack remains damaging: a two-hour hijack with $|P|=20\%$ causes 214 ETH in total damages.

\myitem{Practicality:} This attack is enabled by three key factors—two stemming from Ethereum’s PoS design and one from its Internet-level deployment.
First, an inactivity leak not only halts finality but also imposes financial penalties on both the targeted and unaffected parts of the network, which could provide more incentives for the attacker.
Second, validators are heavily concentrated in a small number of Internet-hosting locations (i.e., ASes), making it feasible for an attacker to isolate them by hijacking just a few IP prefixes. Indeed, we find that 33\% of validators are hosted within only 29 prefixes, while 40.85\% of Ethereum nodes are hosted in prefixes that are not protected by RPKI. 
Critically, BGP hijack attacks involving even thousands of prefixes still happen~\cite{manrs2022bgp}. 
This centralization is not an artifact of our methodology; even applying simpler heuristics, similar to those developed for Bitcoin, yield similar numbers (see appendix ~\ref{apdx:bias}).  
Third, the leak persists regardless of which specific validators are absent. This flexibility allows the attacker to dynamically adjust the targeted validator set $P$, either to compensate for restored connectivity or to correct inaccuracies in IP-to-validator mappings (see \S\ref{sec:stakebleed_selecting_p}).

\subsection{\attacktwo }
\begin{figure*}[h]
    \centering
    \includegraphics[width=0.96\linewidth, trim={0cm 7.45cm 0 0},clip]{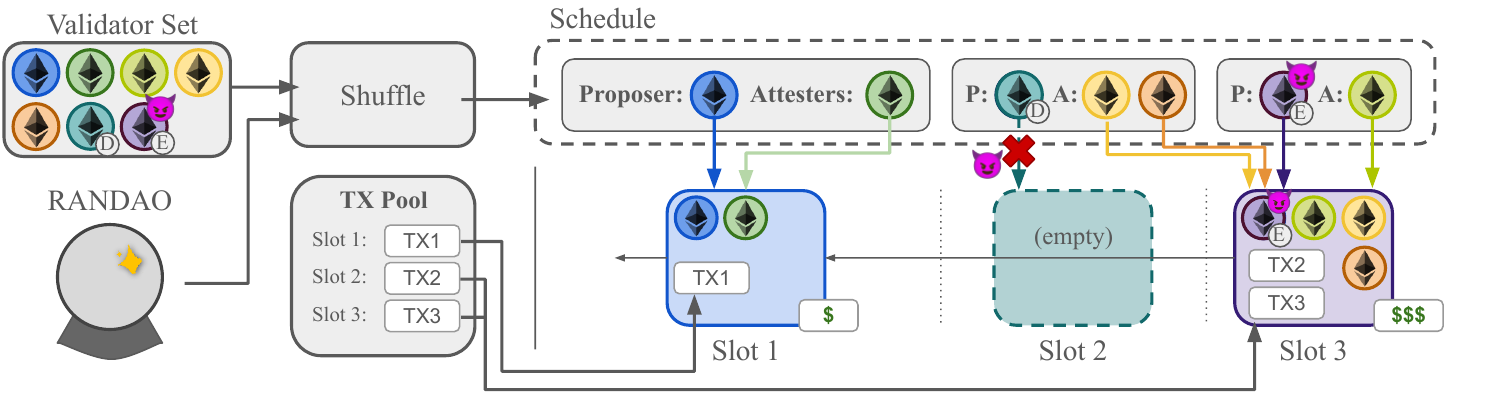}
    \caption{A \attacktwo adversary controls the validator node E that is set to propose in Slot 3. The attacker calculates the proposer schedule, finds that validator D is scheduled to propose before her in Slot 2, and prevents node D from proposing by performing a BGP hijack. She then preferentially includes some transactions and attestations that were intended for Slot 2 in her own block, while denying D its proposer rewards.
    }
    \label{fig:knockblock-overview}
\end{figure*}

In this attack, the adversary described in \S\ref{ssec:threat} seeks to isolate a single, carefully selected validator for a very short period, just enough to prevent it from proposing a block. The attacker owns one validator and seeks to benefit from the Maximal Extractable Value (MEV) expected to be contained in the block. 

\myitem{Description} The attacker first calculates the proposer schedule.
She identifies the validator scheduled to propose a block in the slot immediately preceding her own. Using the mapping obtained via the strategy outlined in \S\ref{sec:identifying_stakeholders}, the attacker launches a BGP hijack targeting the validator’s IP prefix. To increase her success probability the attacker might select the timing of her attack as well as the prefixes to hijack following strategies described in \S\ref{sec:attacktwo_analysis}. Once on-path, she selectively drops all proposer connections for just one slot (12 seconds). Immediately after, she withdraws the BGP advertisement, restoring normal connectivity. Her own validator then proceeds to propose the next block as scheduled, naturally incorporating all unincluded attestations and potentially reusing some of the previously seen transactions to increase her MEV. For an example walk-through of \attacktwo on Fig.~\ref{fig:knockblock-overview}, see \S\ref{apdx:knockblock_example}.

\myitem{Impact} The attack deterministically increases the revenue for the attacker's validator.
We estimate that \attacktwo on average yields 44.5\% higher MEV, and can also lead to an increase in proposer rewards of up to 453.6\%. In addition, the attacker deprives the hijacked proposer of its expected rewards. In \S\ref{sec:attacktwo_analysis}, we show that targeting proposers yields the highest damage per unit of hijack duration. 

\myitem{Practicality} Two factors make this attack practical.
First, the attacker can reliably and promptly predict the proposer schedule up to 2 epochs in advance (12.8 minutes) under stable validator balances~\cite{slot-predictability}. This prediction window not only provides sufficient time to initiate a targeted hijack, but also enables the attacker to strategically select a proposer whose suppression directly benefits her—specifically, the one scheduled to propose immediately before her own validator.
Second, isolating a proposer requires only a brief, single-prefix BGP hijack. Since the proposer only needs to be unreachable for 12 seconds, the attacker’s hijack can be short-lived and indistinguishable from routine Internet misconfigurations~\cite{bgp-misconfiguration}.
Further, since validators are required to operate from a single machine~\cite{redundant-validators} (unlike miners who typically operate distributed infrastructures) the attacker needs to only hijack a single IP prefix, substantially reducing both complexity and risk of detection.

%% file: sections/04_identifying_stakeholders.tex
\section{Identifying Ethereum Stakeholders}
\label{sec:identifying_stakeholders}

Inferring Ethereum’s validator topology requires mapping validators to their nodes (and network locations). Prior methods for identifying PoW miners—based on block propagation timing~\cite{apostolaki2017hijacking,miller2015discovering} or mining pool network information~\cite{tran2024routing}—do not work for validators because validator messages are sent via Gossipsub, which introduces bursty, asynchronous message propagation. Furthermore, connecting \textit{on-demand} to any node is not always possible, and pooled staking interactions happen on the blockchain level, rather than the network level. For more details on these challenges, see \S\ref{apdx:minder_id}.

\subsection{Collecting P2P information}
To overcome these challenges, we design \node, which passively accepts connections and logs validator attestations and auxiliary signals rather than blocks, since attestations are frequent and tied to specific validators.
We explain the challenges this approach entails and the insights that make \node achieve its goal.

\myitem{\node turns attestation timings into a reliable signal.}
Without being connected to all nodes simultaneously, our \node cannot calculate relative timing advantage; that is, our node cannot determine whether one of its peers disseminated an attestation unusually early compared to all other nodes in the network (as it is not connected to all or enough of them).  
Worse yet, because Gossipsub does not force nodes to propagate the attestations they get immediately to all their peers (unlike mined blocks in PoW blockchains), our \node can receive attestations with a substantial delay relative to when they were first known to the senders. 

To address these challenges, we leverage the fact that each validator produces an attestation once per epoch (6.4 minutes) in a fixed 12-second slot. This allows \node to establish relative latency (\texttt{rx\_ts - slot\_ts}) which provides a robust signal for validator identification. \node avoids misleadingly late attestations by classifying incoming attestations and keeping only eagerly pushed ones. Eagerly pushed attestations are those that have not been already advertised via an IHAVE message, and furthermore satisfy one of two conditions: (i) their sender is currently subscribed to the attestation topic and its connection with \node is grafted (mesh), or (ii) they were sent over a non-grafted (fanout) connection and the sender was not subscribed to the topic. We refer to attestations of the latter case as \textit{out-of-scope} (OOS). This method allows us to extract meaningful latency information for every validator for a given peer and compare it across nodes without requiring concurrent global connectivity.

\myitem{\node minimizes per-node connection duration without sacrificing mapping accuracy.}
Unlike PoW mining, where there were only a handful of sources to trace back (5-6 major pools), Ethereum has over one million validators. 
While longer \node connections to peers would in theory give a better signal for validator identification, arbitrarily long connections are not feasible because they waste Ethereum network resources, count against the finite connections that \node can make, and nodes experience significant network identity (IP, node IDs) churn.

We circumvent this using the observation that a node reveals its hosted validators when it eagerly propagates their attestations. Unlike mining pools, where blocks spread through multiple gateways, validators must disseminate their own attestations. Since \node records only eagerly pushed messages, the hosting peer is expected to exhibit the lowest relative latency. Accordingly, \node disconnects and blacklists the peer once it has observed from it at least one attestation for each validator, thereby \textit{finalizing} it. To avoid suppressing attestations from lagging nodes, \node never advertises attestations. 
This adaptive finalization greatly reduces overhead: 60\% of nodes finalize within ~41 hours, while some take much longer (up to 860 - see Fig.~\ref{fig:time2fin}). Without tailoring connection duration, achieving comparable accuracy would require over 20x more resources.

\myitem{\node waits for connections but aggressively grafts them.}
Instead of actively selecting peers, \node passively accepts incoming connections, allowing it to capture nodes that are unreachable due to NATs or restrictive settings. Once a new connection is established, \node grafts its connection to the peer on its subscribed topics to observe eagerly pushed attestations without biasing peer selection. To handle dynamic subscriptions and maximize attestation throughput, \node applies a \textit{graft-on-subscribe} strategy: immediately grafting any new topics for which the peer advertises its subscription. Finalized peers are blacklisted and disconnected.

\subsection{Mapping network entities to validators}

Having finalized all nodes, the goal is to map Ethereum's validator IDs to them. 

\myitem{\node collects multiple in-network signals.}
To map nodes to validators, we leverage several signals, including relative latency, attestation counts, and out-of-scope (OOS) attestations statistics.
Relative latency of received attestations is the strongest signal for three key reasons.
First, validators originate attestations from a single node~\cite{redundant-validators} - hence receiving an attestation through a relaying node will always be slower than learning about it from the source. 
Second, validators are incentivized to publish their attestations as quickly as possible to increase their chances of timely inclusion in a block, and thus acquire their full rewards.
Finally, it is physically impossible to artificially reduce network delay: Ethereum nodes cannot make attestations appear to have arrived any sooner than the sum of gossip communication and processing times.
Attestation and OOS counts provide valuable auxiliary information, but are less reliable on their own, as not all nodes send OOS messages consistently, and counts depend on connection duration.
In particular, OOS attestations should only originate from nodes hosting a validator (as observed by~\cite{heimbach2024deanonymizing}), yet not all such nodes send them—either because they subscribe to all topics (common for nodes with many validators) or the \node is outside the peer's fanout.

\begin{figure*}
    \captionsetup{labelformat=empty}
    \centering
    \begin{subfigure}[t]{0.45\textwidth}
        \centering
        \includegraphics[width=1\linewidth, trim={0 0 0 0.1cm},clip]{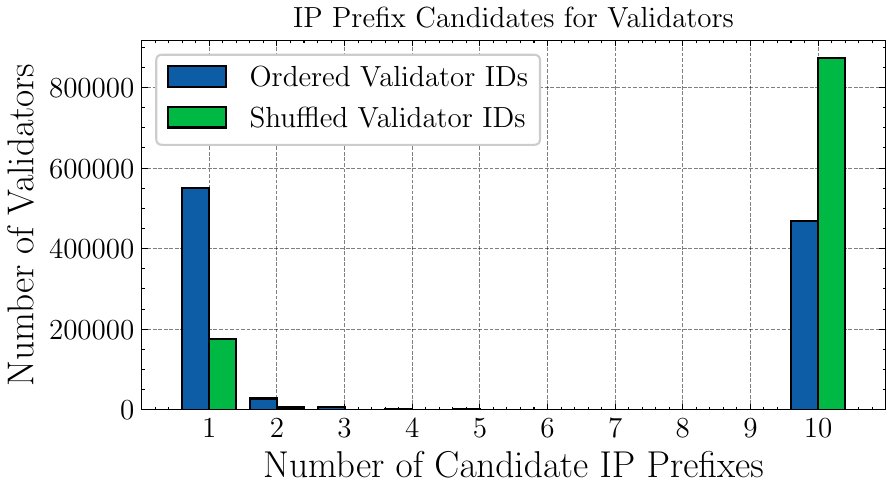}
        \caption{For $>50\%$ of validators, nodes in a single IP prefix have exceptionally low latency in delivering attestations of consecutive validator IDs.
        }
        \label{fig:validx-correlation}
    \end{subfigure}
    \hfill
    \begin{subfigure}[t]{0.47\textwidth}
        \centering
        \includegraphics[width=1\linewidth]{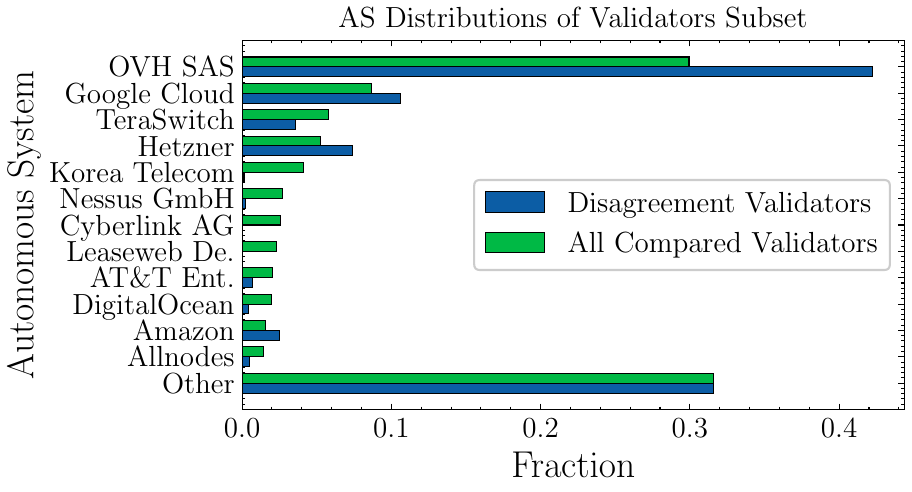}
        \caption{Validators in multi-region cloud environments show reduced mapping consistency, possibly due to IP prefix changes or datacenter migration.
        }
        \label{fig:mismatch_asn}
    \end{subfigure}
    \vspace{-0.4cm}
\end{figure*}

\myitem{\node leverages latency and consecutive validator IDs to map a subset of the validators}
Because a validator’s host is expected to deliver its attestations with lower latency, \node ranks peers by latency for each validator and selects the top 10 candidate IPs. Direct latency comparisons across candidates are unreliable, as measurements occurred at different times for different attestations. Instead, we make the following observation: some validators with consecutive IDs were likely activated in sequence by the same entity staking multiples of 32 ETH at a time, and are hosted under the same ISP, potentially sharing IP prefixes. By computing intersections of candidate node IP prefixes across consecutive validator IDs, we find that 50\% of validators have at least one top candidate node sharing a common prefix. As Figure~\ref{fig:validx-correlation} shows, consecutive IDs typically share either one common prefix or none. Cases with no common prefix may correspond to validators hosted in different locations, no longer active, or hosted on nodes unobserved by our \node. To confirm this mapping is meaningful, we shuffled validator IDs and repeated the analysis, finding almost no common prefixes—demonstrating that our initial mapping is not coincidental or dominated by a few highly connected nodes.

\myitem{\node relies on a small verifiable dataset to generalize inference.}
Although latency is a strong signal, it can be noisy, making manual heuristics insufficient to map all validator-to-node assignments. To leverage complex patterns in timing and auxiliary data, we train a machine learning model to identify which node in a candidate set hosts a given validator. Our intuition is that per-validator statistics for the host node have distinctive validator-specific features compared to other nodes, such as markedly lower latency. The model takes as input the behavior of the ten lowest-latency nodes for a validator and predicts the host. Importantly, we do not expect a learnable pattern in per-node behavior, as this is confounded by factors such as the number of validators a node hosts.

For training, we further filter our previously mapped validators using the intuition that validators sharing the same deposit keys may be co-located in the network. This filtering does not bias the model toward recognizing only co-located validators, as the patterns it learns—e.g., attestation timing, OOS counts—are independent of ownership. The step primarily reduces potential false positives at the cost of some false negatives (\ie mappings that were correct but we are not using them to train).

To implement this, we cluster validators by deposit key and examine their distributions across ASes, scoring each distribution using Efficiency (normalized Shannon Entropy). Near-zero efficiency indicates a skew toward a single AS, i.e., largely homogeneous prefix assignments. While some owners may distribute validators across multiple ASes, we train only on low-efficiency assignments for greater confidence in training quality. Our model is a simple MLP with four hidden layers (3×128, 1×64 neurons) that uses ReLU activations, cross-entropy loss and the Adam (LR=0.001) for 1,000 epochs. To improve generalization, we randomly permute the order of node behavior inputs each epoch. The trained model achieves 91\% classification accuracy, which we use to map all remaining validators to their prefixes.

\myitem{We validate \node's mappings.} 
Validating all mappings is inherently difficult due to the absence of ground truth. However, the lack of common prefixes among shuffled validator IDs and the high accuracy of our ML model increase confidence in the inferred associations. To further support our results, we compare our mappings to those of Heimbach et al.~\cite{heimbach2024deanonymizing}, analyze the hosting environments of Ethereum staking pool operators, and evaluate an alternative mapping methodology.

We compare our mappings against those provided by Heimbach et al.~\cite{heimbach2024deanonymizing}, which, like \node's, could not be readily verified against ground truth.
Their mappings cover approximately 11\% of our mapped validator set, and exclude validators hosted on nodes that subscribe to all 64 attestation gossip topics.
Each validator is associated with a set of candidate countries inferred by up to four independent measurement nodes. Due to the sensitive nature of the data, these mappings do not include finer-grained network identifiers such as ASNs or IP prefixes. Consequently, our comparative analysis may be subject to artifacts introduced by IP-to-geolocation mapping. 
Moreover, the measurements underlying their mappings were conducted in May 2024~\cite{heimbach2024deanonymizing}, whereas \node uses data collected from November 2024 to January 2025. Although large-scale migration is unlikely, some validators may have changed IP prefixes during this interval.

For the comparison, we restrict our analysis to the subset of overlapping validators for which Heimbach et al. have high confidence in their assignment - defined as agreement across all their measurement nodes.
We observe that agreement between the two mappings increases with the confidence of the provided assignments, i.e., with the number of measurement nodes in agreement. For validators with the highest confidence (four agreeing measurement nodes), the two mappings agree for 70\% of validators.
Finally, we analyze the validators where the two mappings disagree. Grouping these validators by the ASN assigned by \node (Figure~\ref{fig:mismatch_asn}), we observe that validators mapped to single-country ASes like Korea Telecom, Nessus, Cyberlink, and Leaseweb are likelier to be assigned to the same country in both methodologies. In contrast, validators hosted on multi-region cloud providers like OVH, Google Cloud, Amazon, and Hetzner are more likely to receive disagreeing country assignments. This pattern could indicate validators in the latter category being reassigned IP prefixes in different geographic locations within the cloud environment, or migration to different datacenters.

We analyze the hosting environments of Ethereum staking pool operators collaborating with Lido~\cite{lido-operators}. Specifically, we identify the ASN hosting each operator’s website and check whether any validators mapped by our methodology fall within the same ASN. For roughly half of the operators, at least one validator is co-located with the website in the same ASN, often corresponding to known IaaS or cloud providers (e.g., Amazon, OVH). For the remaining operators, the ASNs typically belong to CDNs or edge providers (e.g., Cloudflare, Fastly) or security services, which are less likely to host validator infrastructure. This contrast further supports the plausibility of our mappings.

Despite our best efforts, one might still worry that more traditional mapping methodologies would prove our attacks less practical. To investigate this hypothesis, we also generate a mapping using simple heuristics similar to what was used in Bitcoin \eg by \cite{apostolaki2017hijacking}, and report results in appendix \S\ref{apdx:bias}.

%% file: sections/05_topoanalysis.tex
\section{Ethereum PoS Network analysis}
\label{sec:network_analysis}
 
We analyze the Ethereum PoS topology inferred in \S\ref{sec:identifying_stakeholders}, examining validator centralization across IP prefixes, Autonomous Systems (ASes), and geographic locations, which has direct implications for our network-level attacks and Ethereum security.

\myitem{Methodology:}
Using our validator-to-IP mappings, we resolve routable prefixes, ASNs, ROA coverage, and countries. We obtain routable prefixes and ASNs by querying the RouteViews RIB via the pyASN library~\cite{routeviews,pyasn}. Finally, we map countries using IPInfo Lite~\cite{ipinfo-lite}, and check ROA coverage with Routinator~\cite{routinator}.

\begin{figure*}
    \captionsetup{labelformat=empty}
    \centering
    \begin{subfigure}[t]{0.32\textwidth}
        \centering
        \includegraphics[width=\linewidth]{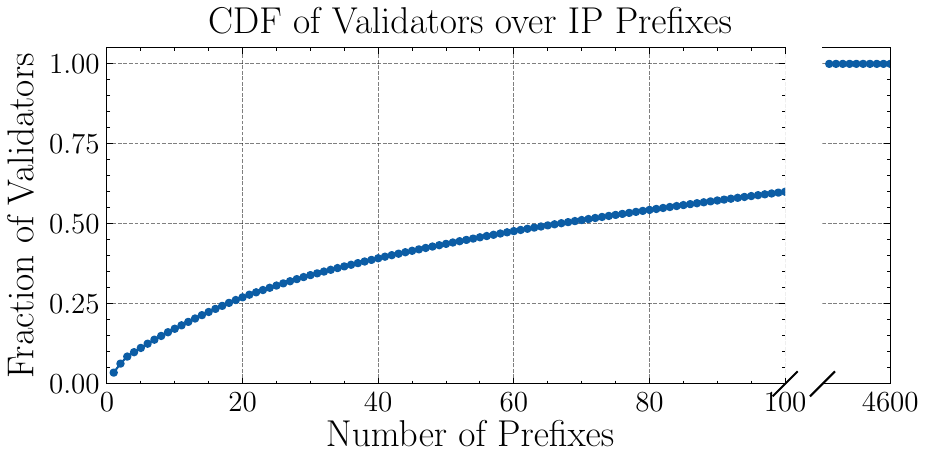}
        \caption{Almost 60\% of validators are hosted in just 100 prefixes.}
        \label{fig:vals-prefix-cdf}
    \end{subfigure}
    \hfill
    \begin{subfigure}[t]{0.32\textwidth}
        \centering
        \includegraphics[width=\linewidth]{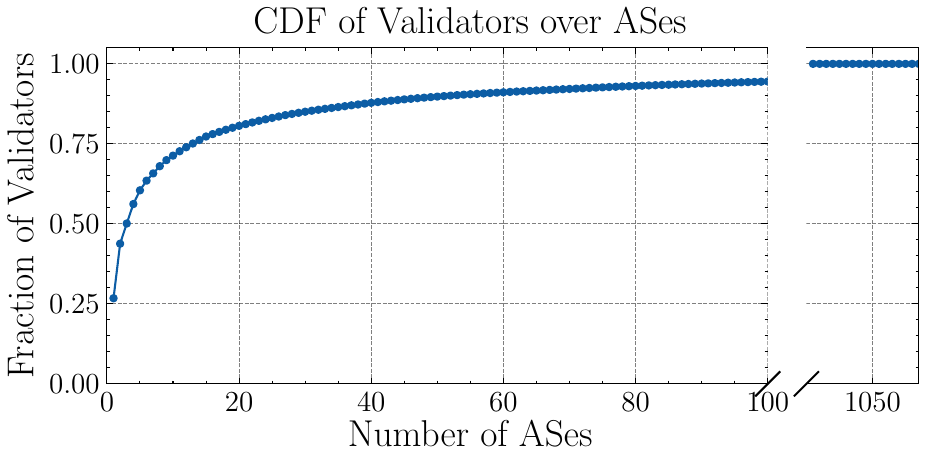}
        \caption{
        Just 3 ASes host $>50$\% of Ethereum's validators.
        }
        \label{fig:vals-asn-cdf}
    \end{subfigure}
    \hfill
    \begin{subfigure}[t]{0.32\textwidth}
        \centering
        \includegraphics[width=0.9\linewidth,trim={0.1cm 0.2cm 0.12cm 0},clip]{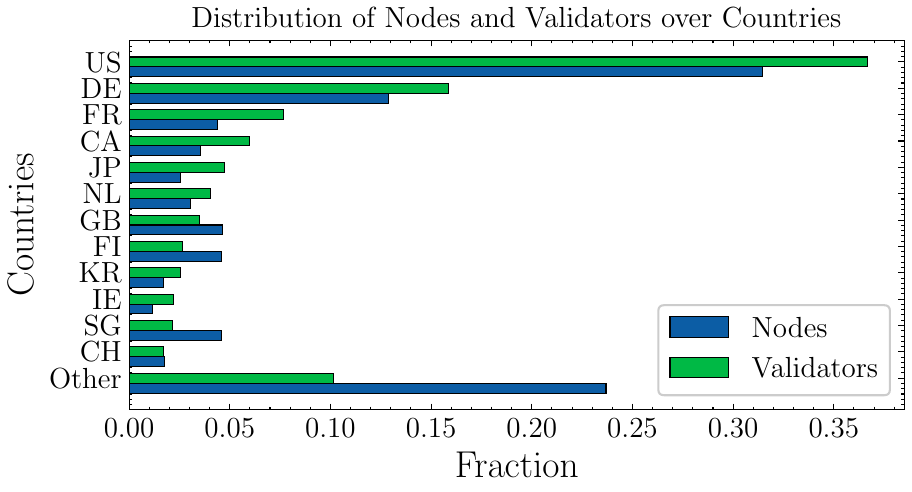}
        \caption{The US hosts $\approx37\%$ of validators and 31\% of nodes.}
        \label{fig:vals-country-dist}
    \end{subfigure}
    \vspace{-0.4cm}
\end{figure*}

\myitem{Validator Centralization} 
High validator concentration increases an attacker’s efficiency in targeting large portions of the network. Ethereum validators are highly centralized at the prefix level: 60\% reside under just 100 prefixes, with individual prefixes hosting up to 37,000 validators (Figure~\ref{fig:vals-prefix-cdf}). The full set ($N=1,063,660$) spans only 4,600 prefixes, and over a third of it is under just 29 prefixes, meaning a StakeBleed attacker could drive the network into the leak state by hijacking just these prefixes. As a robustness check on our centralization estimates, we repeated the analysis using an alternative mapping methodology (see \S\ref{apdx:bias}), which reveals even greater centralization.

Validators are also concentrated at the AS level: just 20 ASes host > 80\% of validators, and the full set spans only 1,057 of 118,000 allocated ASNs (Fig.~\ref{fig:vals-asn-cdf}). Notably, OVH alone hosts almost 27\% of validators, close to the 1/3 threshold needed to disrupt block finalization. 

Geographic centralization is similarly high: approximately 37\% of validators reside in the U.S., and just 12 countries account for nearly 90\% of stake (Figure~\ref{fig:vals-country-dist}). While non-state actors are unlikely to compromise entire countries, a single country or a small coalition could threaten Ethereum consensus by surpassing the 1/3 stake threshold.

%% file: sections/05_attackone.tex
\section{\attackone analysis}
\label{sec:attackone_analysis}

\myitem{Validator rewards} are multiples of the base reward rate $b$ which scales with the inverse of the square root of the total active stake. Considering Ethereum's current validator set size of $N=1,063,660$, we derive $b=346$. Rewards and penalties are applied to the validator's balance, which is used to update its effective balance of 1 ETH increments every epoch. The latter is used in determining the rewards, penalties, and the probability of performing duties other than attesting. In this section we consider all validators to have their initial 32 ETH of actual balance.

Ethereum validators earn on average 11,072 Gwei per epoch (6.4 minutes) in rewards which can be broken down to $R_A$ (9,342 Gwei), $R_P$ (1,384 Gwei/epoch avg.) and $R_Y$ (346 Gwei/epoch avg.) for performing the following duties: casting attestations, proposing blocks, and participating in the sync committee. Validators cast attestations regularly in every epoch, but their selection for block proposal and sync committee participation is random and proportional to their effective balance, creating variance in earned rewards. Attestation rewards can be broken down into $R_A=R_s + R_t + R_h$ (2,422, 4,498 and 2,422 Gwei), for correct and timely votes for what the validator believes to be the \textit{source}, \textit{target}, and \textit{head} blocks respectively. Finally, block proposal rewards are also divided into two parts $R_P = R_{PA} + R_{PY}$ (up to 1,335 and 49 Gwei/epoch avg.), that scale with the number of attestations and sync committee outputs that the proposer includes in the block.

\begin{figure*}
    \captionsetup{labelformat=empty}
    \centering
    \begin{subfigure}[t]{0.32\textwidth}
        \centering
        \includegraphics[width=1\linewidth]{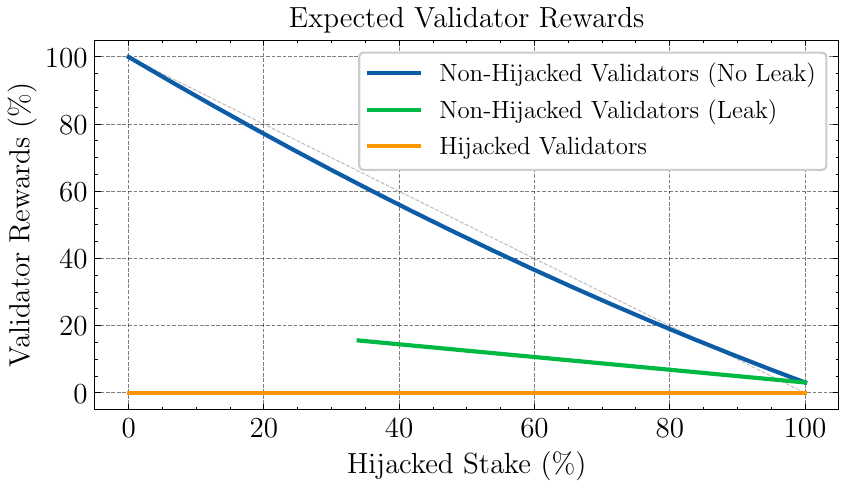}
        \caption{
        Non-hijacked validator rewards decrease as $P$ grows. A leak further reduces them by 84\%. Hijacked validators lose all rewards.
        }
        \label{fig:exp_rewards}
    \end{subfigure}
    \hfill
    \begin{subfigure}[t]{0.32\textwidth}
        \centering
        \includegraphics[width=1\linewidth]{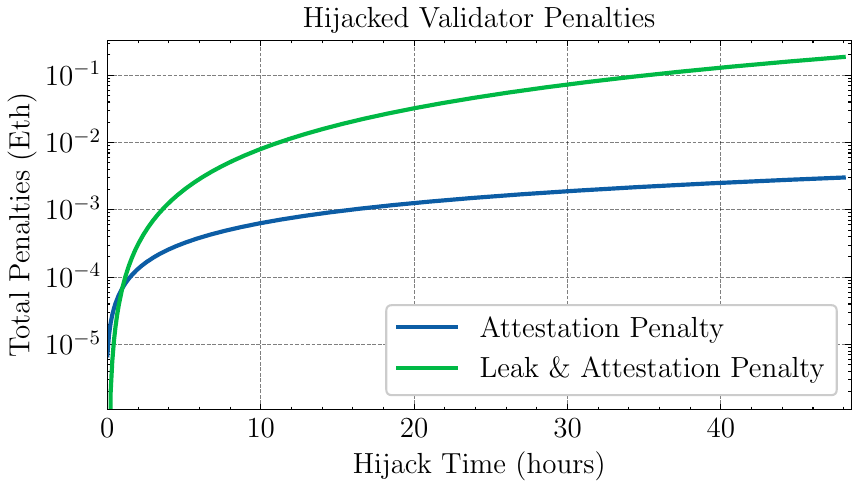}
        \caption{As the attacker maintains the partition, the inactivity penalty grows quadratically 
        over time.
         }
        \label{fig:exp_stake_loss}
    \end{subfigure}
    \hfill
    \begin{subfigure}[t]{0.32\textwidth}

        \centering
        \includegraphics[width=1\linewidth]{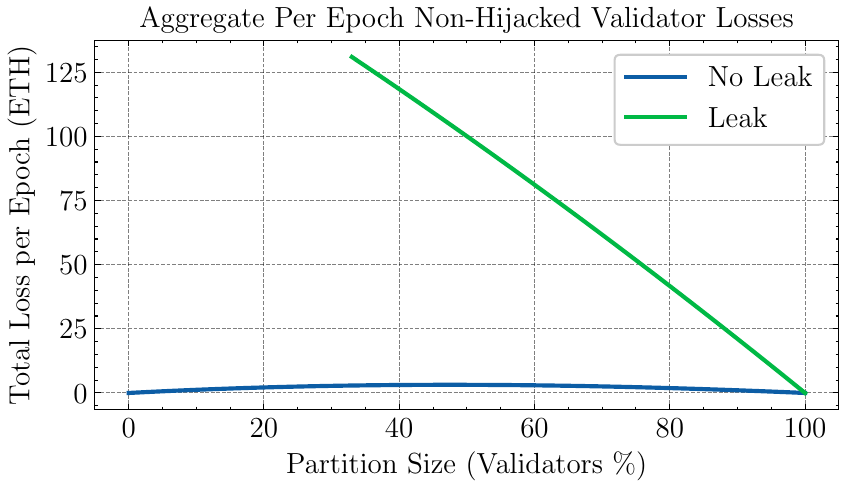}
        \caption{
        The attacker causes two orders of magnitude higher losses to non-hijacked validators when she causes a leak.
        }
        \label{fig:non-hijacked-aggr-losses}
    \end{subfigure}
    \vspace{-0.6cm}
\end{figure*}

\myitem{Non-hijacked validator rewards}: if a \attackone attacker hijacks a $p$ fraction of validators, non-hijacked validators lose on average $13148 \cdot p - 2422 \cdot p^2$ Gwei per epoch in rewards, which increases to $1384\cdot p + 9342$ Gwei if the attacker induces a leak. In the absence of a leak, the losses occur because of three things: (i) attestation rewards as a whole are directly reduced by the fraction of the hijacked stake, (ii) validators voting in a slot of a hijacked proposer are unable to cast a timely head vote, forfeiting that reward, (iii) non-hijacked block proposers are unable to collect attestations and sync contributions from the $p$ fraction of hijacked validators. Putting these together, a validator's expected per-epoch reward becomes $E[R_V] = (1-p)\left[(1-p) R_h + R_s + R_t +  \frac{32\cdot R_P}{N} \right] + \frac{32\cdot512\cdot R_Y}{N}$. If the attacker forces the chain to enter the leak state, $R_h, R_s, R_t$ are set to zero, effectively removing attestation rewards. 
Figure~\ref{fig:exp_rewards} depicts the fraction of the their full expected rewards that validators receive as a function of the hijacked stake. Taken in aggregate, an attacker can cause losses of over 125 ETH per epoch to non-hijacked validators with a partition size of $p>0.33$, as shown in Figure~\ref{fig:non-hijacked-aggr-losses}. Note that all rewards $R_x$ are multiples of the base reward rate $b$. A reduction in total active stake would increase $b$ and consequently all $R_x$ which means that our estimates serve as a lower bound for validator losses.

\myitem{Hijacked validator rewards}: regardless of the leak, hijacked validators miss all rewards since, when they rejoin the network, they adopt the majority view of the chain that non-hijacked validators have as canonical. Instead, hijacked validators incur an attestation penalty that scales as $0.625\cdot n\cdot b$, with $n$ being the validator's effective balance increments of 1 ETH.
For our given $b$, and 32 effective balance increments, this amounts to 6,920 Gwei per epoch. If the attacker triggers a leak, hijacked validators also pay a $\frac{s_i \cdot B_i}{4\cdot1677216}$ Gwei penalty on every epoch, where $s_i$ is the validator's inactivity score at epoch $i$ (increases by 4 per epoch hijacked during the leak), and $B_i$ is the validator's effective balance at epoch $i$ (initially $32\cdot10^9$ Gwei). These penalties are depicted in Figure~\ref{fig:exp_stake_loss} as a function of the hijack length. Depending on the duration of the hijack, the inactivity penalty can continue to apply even after the network exits the leak state. Validators that have their effective stake balance reduced to the 16 ETH threshold are ejected from the active validator set - this takes $\sim$ 3 weeks with a leak for a validator that has a 32 ETH initial balance.

\myitem{Ethereum's blockchain throughput} in Transactions Per Second (TPS) decreases proportionately when an attacker hijacks a fraction $f$ of validators. This is because block proposers are selected pseudorandomly in proportion to their effective balance. Thus, hijacking fraction $f$ of validators prevents the same fraction of block proposals.

\begin{figure*}
    \captionsetup{labelformat=empty}
    \centering
    \begin{subfigure}{0.34\textwidth}
        \centering
        \includegraphics[width=0.95\linewidth]{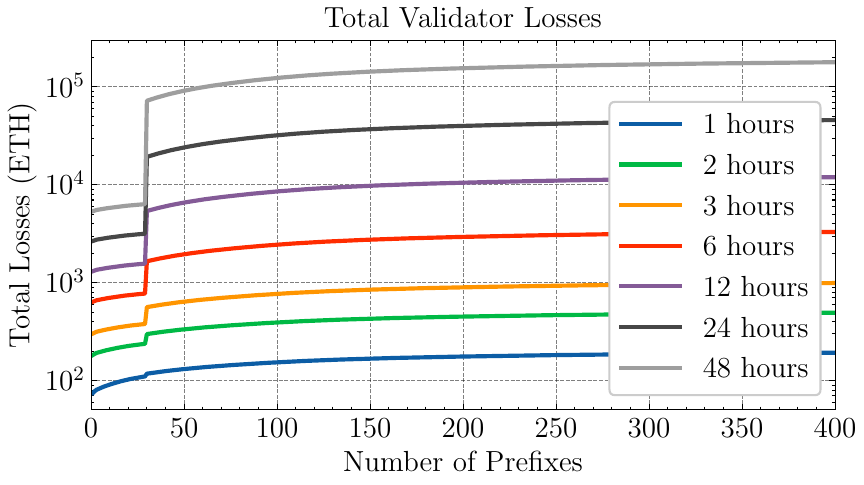}
        \caption{A 30 prefix hijack for 2/3 hours can cause $\sim300/379$ ETH losses with/w.o a leak.}
        \label{fig:prefixes-total-losses}
    \end{subfigure}
    \hfill
    \begin{subfigure}{0.31\textwidth}
        \centering
        \includegraphics[width=1\linewidth]{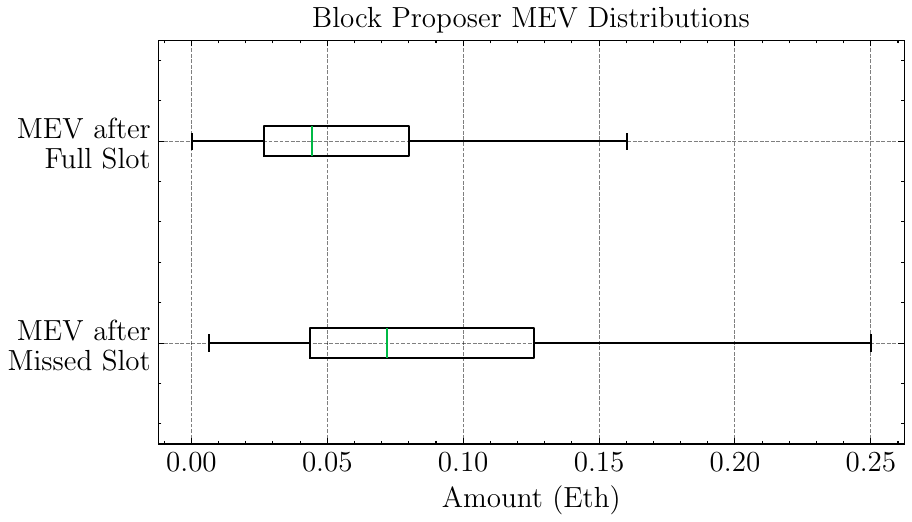}
        \caption{
        Blocks proposed after an empty slot had 44.5\% higher average MEV.}
        \label{fig:block-reward-dist}
    \end{subfigure}
    \hfill
    \begin{subfigure}{0.31\textwidth}
        \centering
        \includegraphics[width=1\linewidth]{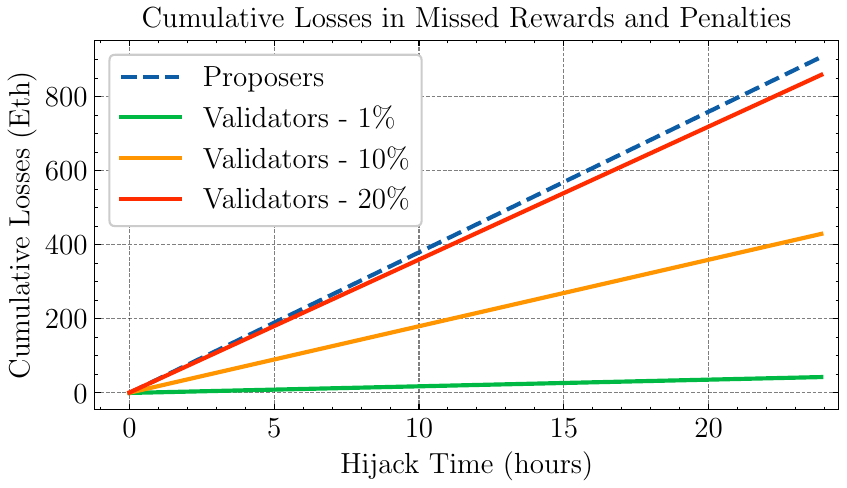}
        \caption{Targeting proposers yields similar damages as hijacking $20\%$ of validators.}
        \label{fig:max-loss}
    \end{subfigure}
    \vspace{-0.4cm}
\end{figure*}

\label{sec:stakebleed-cost}
\myitem{\attackone's costs in practice}:
We estimate the aggregate losses that a realistic adversary can inflict on validators through penalties and missed rewards. Network operators surveyed on BGP hijacks reported the majority of hijacks lasted over an hour, and their effects spanned for hours or more~\cite{sermpezis2018survey}. We thus consider an adversary with a hijacking time budget of a handful of hours during which she prioritizes hijacking prefixes with the most hosted validators to maximize her impact. As shown in Figure ~\ref{fig:prefixes-total-losses}, the attacker significantly improves her ability to inflict losses by crossing the threshold of 29 prefixes required to trigger a leak (see \S\ref{sec:identifying_stakeholders}), causing nearly 300 ETH in damages with just a 2-hour hijack. Our findings also show that even if the attacker does not trigger a leak, prolonging the attack is an effective way to increase her impact, with a 3-hour hijack causing an aggregate of 379 ETH in losses.

\myitem{Experimental validation:}
We experimentally validate \attackone in a simulated Ethereum network. Our simulations take as input the validators distribution, and the changes to the network connectivity graph when the attacker performs a hijack. Our results confirm \attackone's effectiveness, as described in \S\ref{sec:stakebleed-cost} (see \S\ref{apdx:simulation} for details). 

%% file: sections/05_attacktwo.tex
\section{\attacktwo}
\label{sec:attacktwo_analysis}

\myitem{Maximizing attack success:}
To maximize success the attacker targets proposers for which she has high mapping confidence (e.g., those seeded via consecutive validator IDs) and prefixes that are easy to hijack (e.g., prefixes < /24 without an ROA). She may also hijack additional candidate prefixes for the same validator to hedge mapping or filtering failures. If the immediate preceding proposer cannot be suppressed, the attacker can instead target proposers/attesters in the few preceding slots; because blocks include a bounded number of prior-slot attestations, this means the attacker still has a high probability of including attestations intended for previous slots in her block.

\myitem{Proposer predictability:}
\attacktwo requires accurately predicting the proposer schedule with enough lead time for a BGP announcement to take effect. Effective validator balances change only rarely (only 0.0033\% of blocks include a slashing~\cite{rated_network}), and the RANDAO-based schedule is available two epochs in advance, giving an attacker a practical prediction window of $\approx$12.8 minutes. BGP announcements can take effect in less than 90 seconds~\cite{apostolaki2017hijacking}, so the attacker can reliably prepare short, targeted hijacks.

\myitem{Impact analysis:}
\label{sec:knockblock_impact}
Successful KnockBlock attacks lead to a measurable increase of up to 453.6\% in the attacker's revenue in the form of block proposal rewards because of the inclusion of additional attestations and sync committee contributions. Using the protocol specification the expected block proposal reward is $E[R_P] = \frac{1}{7} \cdot \left[ \frac{N}{32} \cdot R_A + 512 \cdot R_Y \right]$.
In the event that there exist valid attestations issued in the preceding 32 slots that have not been included in another block, the block proposer can include these for an additional reward. In particular, new attestations receive the full reward of $\frac{54\cdot n \cdot b}{64}$ Gwei. Attestations delayed up to 5 slots lose the head portion of their reward, meaning that they receive $\frac{40\cdot n \cdot b}{64}$ Gwei. Attestations delayed 6 to 32 slots also lose the source portion of the reward and only receive the target portion: $\frac{26\cdot n \cdot b}{64}$ Gwei. A proposer can include up to 8 slots worth of attestations~\cite{eip-7549} in their block, increasing their reward by up to 453.6\%.
A single block proposal is worth 46,003,328 Gwei, while a block proposed by a \attacktwo adversary is worth up to 254,674,423 Gwei.
Using mevboost payment data~\cite{mevboost.pics}, we observe that blocks following a missed slot yield, on average, 44.5\% higher MEV (Fig.~\ref{fig:block-reward-dist}). As a result, by forcing a proposer to miss its slot, a \attacktwo adversary also increases her expected MEV by 44.5\%.

\myitem{Attack costs:} A \attacktwo adversary can use her profits from running the attack to amortize her costs in setting up her own AS (RIR fees~\cite{APNIC-fees}, transport providers~\cite{own-asn}, hardware~\cite{nick-isp}, etc.) if she does not run or has already compromised one. Given that a validator proposes on average $\sim2.47$ times per year, and that \attacktwo attack results in an average increase of 0.05 ETH in MEV and 0.03 ETH in attestation rewards, the validator's annual revenue increase because of the attack will be
0.19 ETH (848 USD at the time of writing). By estimating annual AS operating costs to be approximately \$2,500, the attack is profitable if the adversary owns more than 3 validators. This is a very pessimistic estimate for the attacker, considering that she can increase her revenue at no additional cost by collaborating with other validator owners and targeting the most profitable MEV opportunities.

\myitem{Comparative impact} 
A focused \attacktwo campaign can cause aggregate network damage comparable to larger \attackone partitions while being substantially stealthier (Fig.~\ref{fig:max-loss}). For example, with a 24-hour active hijack budget, both attack types can impose >800\,ETH in losses, but \attacktwo achieves similar damage via many brief, single-prefix hijacks rather than sustained multi-prefix hijacks.

\label{sec:knock_atk_eval}

%% file: sections/09_countermeasures.tex
\section{Countermeasures}
\label{sec:countermeasures}
\myitem{Automate migration on local detection:} 
A key challenge in mitigating BGP hijacking is that its impact cannot be addressed at the application layer, nor can it be resolved locally. Recovery requires intervention by the affected network provider, who must coordinate with upstream and peer providers to filter the bogus advertisement \emph{at their networks}—a process that often takes several hours~\cite{cloudflare2021facebook,cloudflare2024incident}.
This makes it essential for validator operators to implement migration strategies, meaning automated processes that relocate validator operations to alternative (cloud) locations, under different providers and separate jurisdictions to minimize shared dependencies. This can be triggered by observable signals, such as changes in peer connectivity, unexpected reward loss, or increased message delays, which might indicate malicious reroutes~\cite{milolidakis2023effectiveness}.

\myitem{Reducing predictability:} 
\attackone is feasible because the attacker can monitor validator activity and adapt the partition accordingly. Knowing when each validator is expected to attest, she can verify whether she has successfully isolated them.  \attacktwo relies on the adversary's knowledge the block proposer schedule in advance, reducing the BGP hijacking set to even a single prefix. 
Removing the predictability in both cases would significantly strengthen the protocol against such attacks. For hiding the proposer schedule, one can use one of the several efficient SSLE constructions~\cite{catalano2023efficient, boneh2020single}. In fact, integrating SSLE into Ethereum is actively being discussed~\cite{whisk}.

\myitem{Obfuscating validator locations:} 
Our attacks depend on accurately mapping validators to IP addresses using P2P timing signals. We found that naive countermeasures are ineffective; adding exponentially distributed latency ($1/\lambda=[50, 500]$ms), for instance, barely reduces our mapping model's accuracy (1-3\%) and risks harming liveness. This finding is consistent with prior work in Bitcoin~\cite{10.1145/3224424}. A more effective approach is to disrupt the identifying signals at the source. Disabling "eager push" and OOS attestations is highly effective, slashing our inference model's accuracy from 93\% to 38\%. Alternatively, validators could use temporary proxy nodes to broadcast attestations, effectively serving as decoys to obscure their true location.

%% file: sections/11_related_work.tex
\section{Related Work}
\label{sec:related_work}
\myitem{Network attacks and defenses on PoW} Routing attacks on cryptocurrencies have previously been discussed in the context of partitioning the Bitcoin network~\cite{apostolaki2017hijacking,saad2019partitioning}. Tran et al. demonstrated that it is possible to attack PoW mining pools by tampering with Stratum connections intercepted via BGP hijacks~\cite{tran2024routing}. 
Researchers have also attempted to partition multiple cryptocurrencies at once by inferring Bitcoin's and Ethereum's network topologies through Ripple's~\cite{threebirds2023}. Neither of these works leverages a PoS-specific mechanism, nor do they infer the Ethereum PoS topology.  
Eclipsing attacks on blockchain peer-to-peer networks~\cite{HKZG_15,heo2023partitioning,tran2020stealthier,saad2021syncattack},
are not routing attacks per se as they do not perform a BGP hijack and can be resolved with a single whitelisted connection.
Defenses against routing attacks include SABRE~\cite{sabre}, which, however, contributes to the network's consolidation and offers little incentive for newcomers.

\myitem{Attacks on Ethereum PoS consensus} Recent research in Ethereum PoS has focused on manipulating its fork choice rule and finality gadget to degrade the blockchain's performance, yet without exploiting network routing. 
These methods allow an attacker to perform reorgs~\cite{neuder2021low,three-attacks-on-pos}, and split validators across different chains~\cite{neu2021ebb}.
In response, Ethereum introduced countermeasures and patches, such as proposer boost, but these have been shown to still be vulnerable~\cite{two-more-attacks, pos-under-scrutiny}. Finally, prior work has investigated ways to further exploit Ethereum's inactivity leak~\cite{10646904} after it has been triggered if the attacker holds stake. Such attacks can be used atop \attackone to increase its impact.

%% file: sections/12_conclusion.tex
\section{Conclusion}
This paper presents the first comprehensive analysis of Ethereum’s PoS from the Internet routing perspective, making two key contributions. First, we uncover a new class of network-layer attacks that exploit routing vulnerabilities and PoS mechanisms. We demonstrate two representative attacks: \attackone, which can halt finality and causes losses of almost 300 ETH in 2 hours by hijacking as few as 30 prefixes, and \attacktwo, which can boost MEV gains by 44.5\% with a hijack lasting under 2 minutes. Second, we design a novel \node that scalably logs timing information with minimal disruption to the live network, and introduce a multistage algorithm to infer the Internet-level distribution of validators from this timing data. We also propose a set of readily deployable countermeasures that move beyond idealistic calls to secure the entire Internet and focus on practical, actionable, adopter-facing defenses.

%% file: sections/15_acks.tex
\section{Acknowledgments}
This work was supported by the Princeton Innovation Fund Award and the Princeton DeCenter. We thank the authors of Heimbach et al.~\cite{heimbach2024deanonymizing} for sharing their mappings, which enabled our comparison.

%% file: sections/14_appendix.tex
\appendix

\section{\attackone Example}
\label{apdx:stakebleed_example}
We illustrate the \attackone attack in Fig.~\ref{fig:stakebleed-overview} where the left figure corresponds to the Internet level view and the right to the blockchain view.
We assume a small representative network of 8 ASes (Fig.~\ref{fig:stakebleed-overview}), with some of these hosting Ethereum nodes with validators. The adversary controls a BGP-speaking router in AS 8 and aims to isolate the shaded nodes on the left side of the network. To achieve this, she advertises IP prefixes that cover the IP addresses of Ethereum clients A, B, and C. As a result, she intercepts traffic destined for these nodes, effectively enabling her to drop all connections between nodes in the target set $P$ and the rest of the network. For example, the attacker drops traffic between nodes A and G, as illustrated by the red dashed lines.
The results of the attack at the blockchain level are shown in Fig.~\ref{fig:stakebleed-overview}. If the attack had not taken place, blocks 1, 2 and 3 would have been proposed by nodes A, B, and F, respectively and attested by a subset of the nodes in the network. During the attack nodes in $P$ and outside it have distinct views of the chain. From the perspective of nodes outside of $P$, which will eventually prevail (canonical chain), node B appears to have failed to propose block 2 when it was supposed to. As a result, it is deprived of its proposer rewards. From the perspective of the canonical chain, the next block (block 3) will be proposed normally as node F is not in $P$, although some attestations will be missing because the corresponding validators are isolated reducing F's rewards from including attestation rewards. Nodes in $P$ set to attest in that block will see penalties (illustrated with red dollars). 
Four epochs (or 25.6 minutes) later, the chain enters the leak state; meaning that the validators on nodes A, B, and C are considered inactive by those on D, E, F, and G, and thus incur both the attestation and inactivity penalties. At the same time, nodes D, E, F and G stop earning attestation rewards and only earn rewards when they propose blocks or contribute to the sync committee.

\section{\attacktwo Example}
\label{apdx:knockblock_example}
We illustrate the \attacktwo attack in Fig.\ref{fig:knockblock-overview}, where the adversary's validator node E is set to propose at Slot 3. The adversary calculates the proposer schedule ahead of time and identifies the validator ID that is scheduled to propose at Slot 2, namely that belonging to node D. By preventing node D from proposing during Slot 2, the attacker extends the pool of available transactions her validator can include and gains additional rewards from unincluded attestations. In the meantime, the validator of node D loses their proposer rewards. 

\section{Miner identification vs Validator identification}
\label{apdx:minder_id}
\begin{figure*}
    \captionsetup{labelformat=empty}
    \centering
    \begin{subfigure}{0.5\textwidth}
        \centering
        \includegraphics[width=1\linewidth]{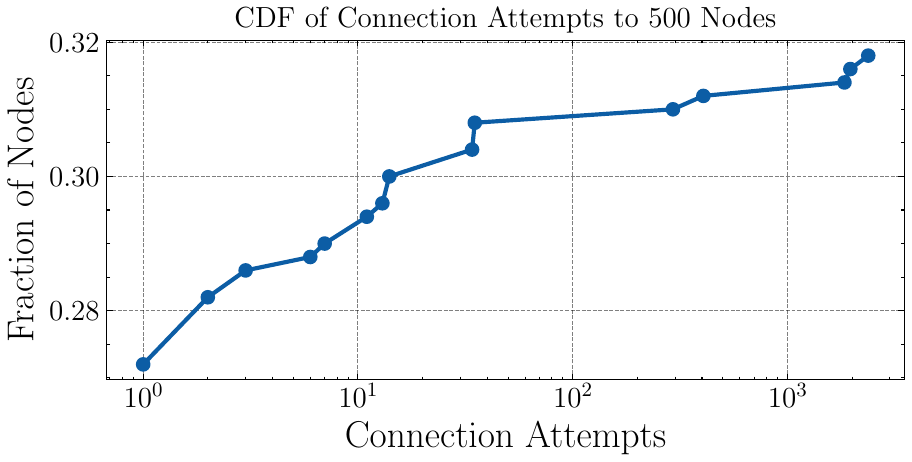}
        \caption{Out of 500 randomly selected node ENRs, our node succeeds in connecting to $<1/3$ them on demand.
        }
        \label{fig:connections}
    \end{subfigure}
    \hfill
    \begin{subfigure}{0.44\textwidth}
        \centering
        \includegraphics[width=1\linewidth]{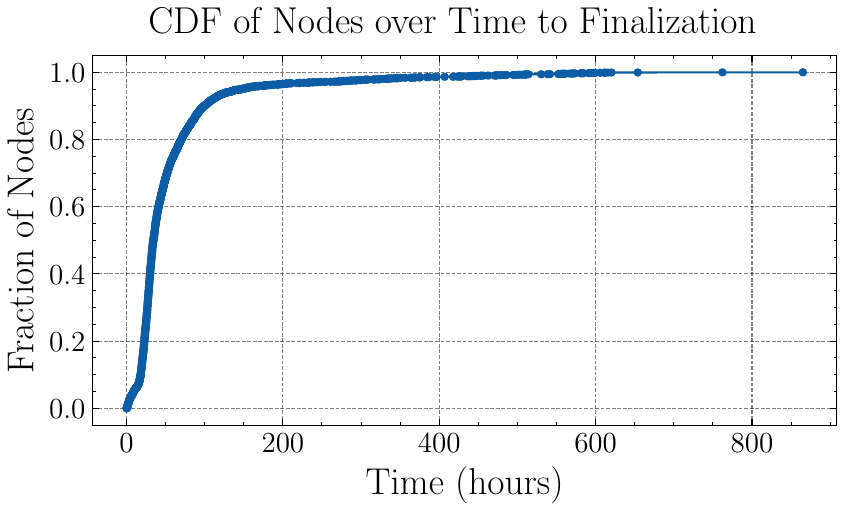}
        \caption{CDF of \node's node finalization time. 60\% of node finalizations happened in just over 41 hours.
         }
        \label{fig:time2fin}
    \end{subfigure}
    \hfill
\end{figure*}

The first approach used to identify miners involves deploying multiple supernodes, namely real Bitcoin/Ethereum clients with multiple peers, which would log the timestamp when each of their peers advertised each block~\cite{apostolaki2017hijacking,miller2015discovering}. 
By analyzing the relative delays in advertising each block across different nodes, one could potentially identify the source of a block. This method, however, relies on two key assumptions: \emph{(i)} that it is possible to maintain concurrent connections with a significant portion of the network to calculate relative delays, and \emph{(ii)} that clients eagerly advertise blocks to all their peers. 
Unfortunately, neither holds true in Ethereum today. 
Because of the extremely bursty communication pattern of the Gossipsub protocol, maintaining connections to all nodes in the network (say 9,000) would require receiving, processing and logging information through the Internet at the impractical rate of 216 Gbps.\footnote{This considers a low-estimate of 24 Mbps per node, which is what we measured in practice but much lower than the theoretic average of a fully-subscribed node.} While this is not the average rate required, inability to keep up with incoming bursts will cause nodes to disconnect and unreliable timestamps. 
Beyond the scaling issue, we find that connecting to that many nodes on demand is impossible. Fig.~\ref{fig:connections} demonstrates our attempt to connect to nodes corresponding to 500 ENRs. Even after 2500 attempts and 625 hours of trying (attempting every 15 minutes), 68\% of the nodes are still not connected. 
Additionally, Ethereum nodes do not immediately propagate messages to all their peers; instead, they only do so selectively (\S\ref{sec:identifying_stakeholders}). 

More recent approaches have relied on the public presence of mining pools and the predictability or advertisement of their stratum server URLs. In Proof-of-Work (PoW), independent mining is highly risky since rewards are only earned by successfully mining a block. Mining pools openly advertise their stratum servers, enabling miners to contribute their computational power. As a result, public mining pools must (directly or indirectly) disclose this information to attract participants~\cite{tran2024routing}. 
However, this is not the case for Ethereum’s Proof-of-Stake system. Anyone with 32 ETH can independently and securely be a validator, eliminating the need for public disclosure or reliance on vulnerable protocols. Even pooled stakers do not usually interact with their respective pools on the network level, but rather do so indirectly through on-chain smart contracts~\cite{pooled-staking}.

\section{An alternative assignment strategy}
\label{apdx:bias}



To validate the accuracy of our findings in Section~\ref{sec:network_analysis}, and assert that they are not negatively affected by the use of our mapping heuristics related to validator management (i.e., using consecutive validator IDs, and low-efficiency validators), we repeat the mapping process using an alternative heuristic that deliberately avoids considering validator ownership. Concretely, we consider the top 10 candidate prefixes for each validator by their relative latencies, and then select the candidate with the highest number of out-of-scope (OOS) attestations. 

\begin{figure}
    \centering
    \includegraphics[width=0.57\linewidth]{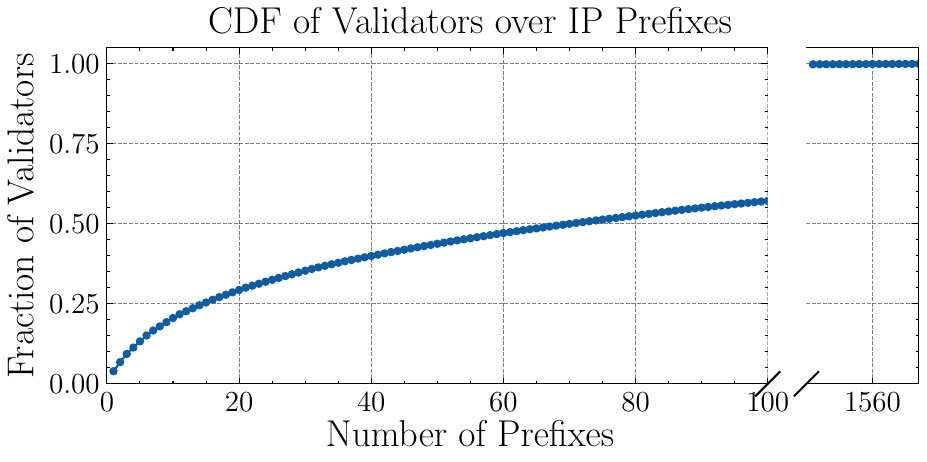}
    \caption{Distribution of validators under prefixes using our simplified heuristic. Almost 60\% of validators are hosted in just 100 prefixes.}
    \label{fig:vals-prefix-cdf-alt}
\end{figure}

\begin{figure}
    \centering
    \includegraphics[width=0.57\linewidth]{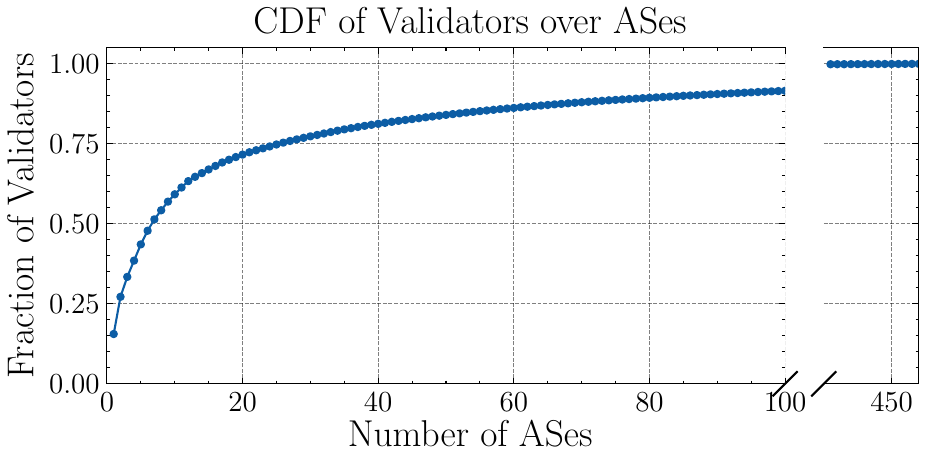}
    \caption{
    Distribution of validators under ASes using our simplified heuristic. Just 100 ASes host almost $90$\% of Ethereum's validators.
    }
    \label{fig:vals-asn-cdf-alt}
\end{figure}

We compare the findings from this simplified mapping heuristic to those by our method in Section~\ref{sec:network_analysis}, and find them to be on par. As illustrated in Fig.~\ref{fig:vals-asn-cdf-alt}, just 20 ASes host nearly 75\% of Ethereum's validators, while Fig.~\ref{fig:vals-prefix-cdf-alt} shows that nearly 60\% of all validators are concentrated in 100 prefixes. These observations are similar to those made in Section~\ref{sec:network_analysis}. Where the simplified heuristic differs from our method in Section~\ref{sec:identifying_stakeholders}, is in the number of ASes and prefixes to which it makes validator assignments. Results from the simplified heuristic suggest a higher network centralization than that indicated in Section~\ref{sec:identifying_stakeholders}.

\section{Simulation Setup}
\label{apdx:simulation}
\myitem{Simulation:}
Existing solutions for instantiating virtual devnets~\cite{ethshadow, jansen2011shadow,du2022seed} either do not support simulating a given validator distribution or only offer limited configuration options for simulating our attacks. We adopt Ethereum Package~\cite{ethereum-package}, a package built by the Ethereum Foundation DevOps team for the Kurtosis~\cite{kurtosis} platform. We then modify Kurtosis to enable fine-grain networking control of the nodes, and write our own orchestration layer that uses a combination of firewall and traffic control rules to emulate the effects of a BGP hijack given its connectivity graphs.

\begin{figure*}
    \captionsetup{labelformat=empty}
    \centering
    \begin{subfigure}{0.60\textwidth}
        \centering
        \includegraphics[width=0.9\linewidth]{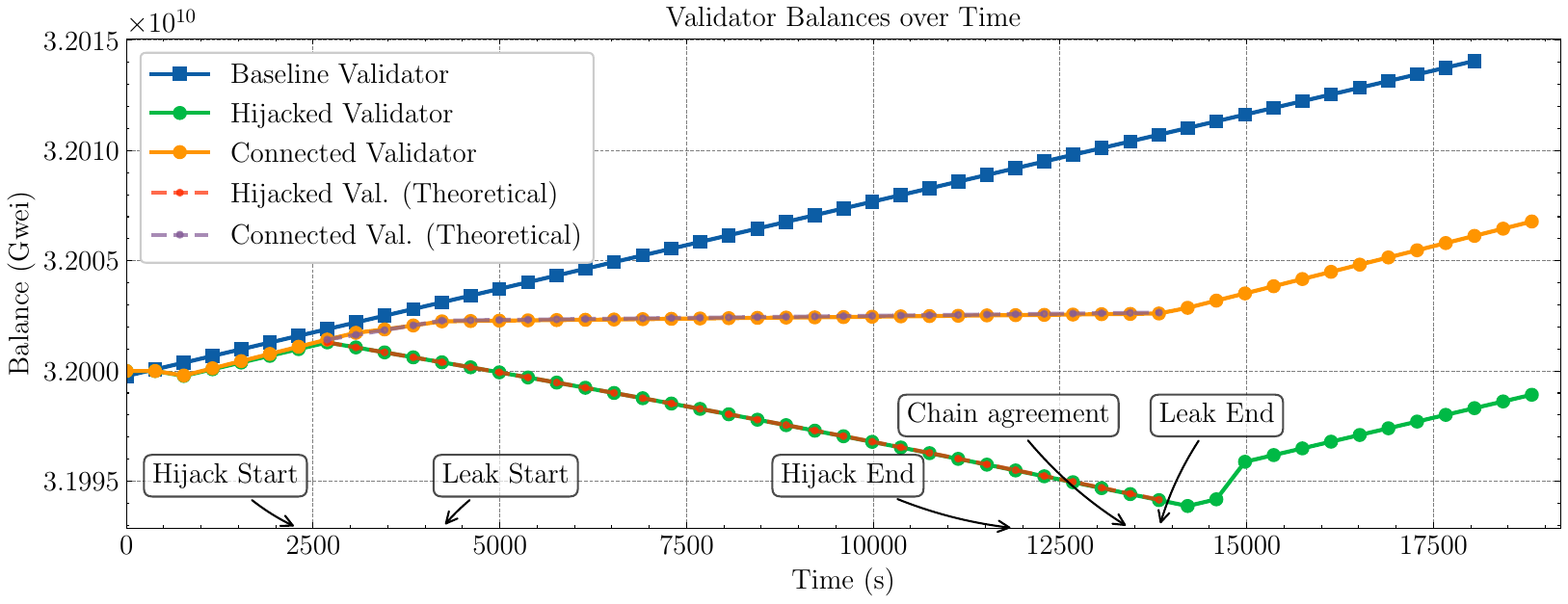}
        \caption{Baseline validator balances are recorded without adversarial activity. In the adversarial setup, a hijack triggers a leak 4 epochs after its start. When the hijack stops, the hijacked validator node regains connectivity, and eventually follows the canonical chain.}
        \label{fig:leak-timeline}
    \end{subfigure}
    \hfill
    \begin{subfigure}{0.36\textwidth}
        \centering
        \includegraphics[width=1\linewidth]{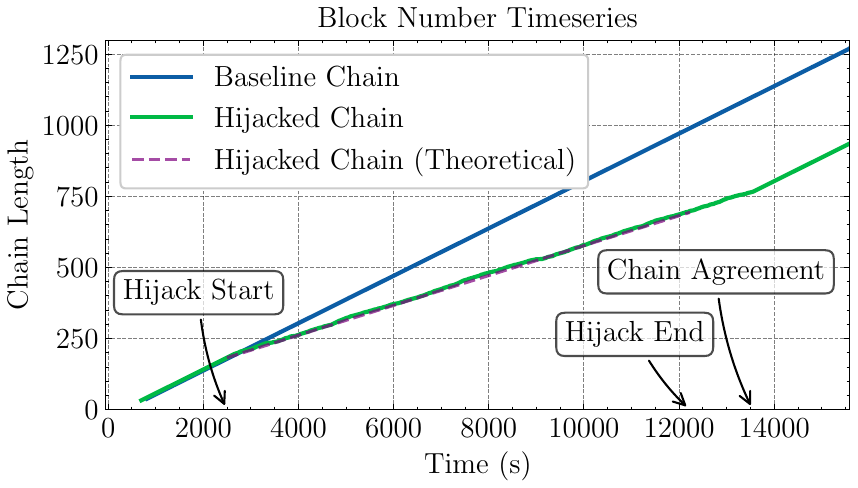}
        \caption{Ethereum block chain growth slows down proportionately to the percentage of hijacked validators. \\}
        \label{fig:chain-length}
    \end{subfigure}
    \hfill
    \caption{}
\end{figure*}

\myitem{Attack Scenario:}
We start the devnet with a strongly connected topology of 20 full Ethereum nodes, the maximum number of nodes that Kurtosis allows us to run on our server. We then assign validators to our devnet nodes by sampling the validator distribution derived in Section \ref{sec:identifying_stakeholders}, and get a devnet of $N=1,008$ validators.
The attacker chooses to create a 37\% partition by hijacking the connections to 3 of the nodes. The attacker initiates the hijack at second 2,496 of the simulation (Fig.~\ref{fig:leak-timeline}), and resolves the partition at second 12,228.

\myitem{Observed Effects:}
True to the protocol, the simulation shows that the network enters the leak state after failing to finalize a checkpoint for 4 consecutive epochs. As can be seen in Figure~\ref{fig:leak-timeline}, connected and hijacked validators match their theoretically calculated balances. Connected validators initially have their rewards reduced proportionately to the hijacked partition size, and after the leak starts, they lose their attestation rewards entirely. Hijacked validators incur attestation and inactivity penalties, and even after the hijack ends and their nodes agree to follow the canonical chain, they continue to incur the inactivity penalty until their inactivity score eventually reaches zero. As also shown in Figure~\ref{fig:chain-length}, we confirm that the chain throughput is reduced according to the hijacked partition size, since approximately 37\% of slots are missed. Note that the absolute values of penalties and rewards in our simulation differ from those of the Ethereum mainnet due to our simulation having a higher $b$ because it hosts fewer validators.

\input{sections/13_ethics}

%% file: sections/13_ethics.tex
\section{Ethical Considerations}
\label{sec:ethics}
We discuss the ethical implications of our work using the Menlo report framework~\cite{bailey2012menlo}. With regard to \textbf{respect for persons}, our work does not directly involve physical persons or their identities, but rather validator entities which we map to their network information. Considering \textbf{beneficience}, our mappings can be used to mount the proposed or other attacks against validators; thus, we intend to make the code and data available only upon request for review purposes or upon approval of the affected parties. We further intend to publish a high-fidelity yet anonymized version of the produced dataset for further research.  To protect validators, we also discuss countermeasures to our methods. In the spirit of \textbf{justice}, our work is relevant to all validators, and benefits the entire Ethereum ecosystem. In \textbf{respect for public law and interest}, our work does not exploit any vulnerabilities (including performing BGP hijacks), uses only publicly available information, and aims to improve Ethereum's security posture.

%% file: bibliography.bib
@article{klayswap_bgp,
	title        = {KlaySwap – Another BGP Hijack Targeting Crypto Wallets},
	author       = {Siddiqui, Aftab},
	year         = 2022,
	journal      = {MANRS},
	url          = {https://manrs.org/2022/02/klayswap-another-bgp-hijack-targeting-crypto-wallets/},
	urldate      = {2024-08-16},
	note         = {accessed 16 August 2024}
}

@article{celerbridge_bgp,
	title        = {How 3 hours of inaction from Amazon cost cryptocurrency holders \$235,000},
	author       = {Goodin, Dan},
	year         = 2022,
	journal      = {Ars Technica},
	url          = {https://arstechnica.com/information-technology/2022/09/how-3-hours-of-inaction-from-amazon-cost-cryptocurrency-holders-235000/},
	urldate      = {2024-08-16},
	note         = {accessed 16 August 2024}
}

@inproceedings{apostolaki2017hijacking,
	title        = {Hijacking bitcoin: Routing attacks on cryptocurrencies},
	author       = {Apostolaki, Maria and Zohar, Aviv and Vanbever, Laurent},
	year         = 2017,
	booktitle    = {2017 IEEE symposium on security and privacy (SP)},
	pages        = {375--392},
	organization = {IEEE}
}

@inproceedings{saad2021syncattack,
	title        = {Syncattack: Double-spending in bitcoin without mining power},
	author       = {Saad, Muhammad and Chen, Songqing and Mohaisen, David},
	year         = 2021,
	booktitle    = {Proceedings of the 2021 ACM SIGSAC conference on computer and communications security},
	pages        = {1668--1685}
}

@inproceedings{threebirds2023,
	title        = {Three Birds with One Stone: Efficient Partitioning Attacks on Interdependent Cryptocurrency Networks},
	author       = {Saad, Muhammad and Mohaisen, David},
	year         = 2023,
	booktitle    = {2023 IEEE Symposium on Security and Privacy (SP)},
	pages        = {111--125},
	doi          = {10.1109/SP46215.2023.10179456}
}

@inproceedings{tran2024routing,
	title        = {Routing Attacks on Cryptocurrency Mining Pools},
	author       = {Tran, Muoi and von Arx, Theo and Vanbever, Laurent},
	year         = 2024,
	booktitle    = {2024 IEEE Symposium on Security and Privacy (SP)},
	pages        = {190--190},
	organization = {IEEE Computer Society}
}

@inproceedings{du2022seed,
	title        = {SEED emulator: An Internet Emulator for research and education},
	author       = {Du, Wenliang and Zeng, Honghao and Won, Kyungrok},
	year         = 2022,
	booktitle    = {Proceedings of the 21st ACM workshop on hot topics in networks},
	pages        = {101--107}
}

@article{ethshadow,
	title        = {Ethshadow: Discrete-event Ethereum network simulator},
	author       = {Knopik, Daniel},
	year         = 2024,
	journal      = {Ethereum Protocol Fellowship - Cohort 5},
	url          = {https://github.com/ethereum/ethshadow}
}

@misc{ethereum-package,
	title        = {Ethereum Package},
	author       = {{Ethereum Foundation DevOps Team}},
	year         = 2022,
	url          = {https://github.com/ethpandaops/ethereum-package}
}

@article{jansen2011shadow,
	title        = {Shadow: Running Tor in a box for accurate and efficient experimentation},
	author       = {Jansen, Rob and Hopper, Nicholas J},
	year         = 2011
}

@misc{kurtosis,
	url          = {https://github.com/kurtosis-tech/kurtosis},
	note         = {Kurtosis: A platform for packaging and launching ephemeral backend stacks with a focus on approachability for the average developer},
	key          = {Kurtosis}
}

@article{eip-7549,
	title        = {{EIP-7549: Move committee index outside of the signed Attestation message}},
	author       = {Dapplion and Kalinin, Mikhail},
	year         = 2023,
	journal      = {Ethereum Improvement Proposals},
	url          = {https://eips.ethereum.org/EIPS/eip-7549}
}

@misc{slot-predictability,
	title        = {{EIP-7917: How often does it help?}},
	author       = {Dapplion},
	year         = 2025,
	url          = {https://ethereum-magicians.org/t/eip-7917-deterministic-proposer-lookahead/23259},
	note         = {Accessed April 11 2025}
}

@misc{routeviews,
	title        = {{Route Views Project}},
	author       = {{The University of Oregon}},
	url          = {http://www.routeviews.org/routeviews/}
}

@misc{routinator,
	title        = {{Routinator}},
	author       = {{NLnet Labs}},
	url          = {https://www.nlnetlabs.nl/projects/routing/routinator/}
}

@misc{rated_network,
	title        = {Rated Explorer},
	author       = {Rated.Network},
	url          = {https://explorer.rated.network/slashings?network=mainnet&timesSlashedPage=1&slashesReportedPage=1}
}

@misc{redundant-validators,
	title        = {Staking Launchpad: Validator Checklist},
	author       = {{Ethereum Foundation}},
	url          = {https://launchpad.ethereum.org/en/checklist},
	note         = {Accessed April 18 2025}
}

@article{sermpezis2018survey,
	title        = {A survey among network operators on BGP prefix hijacking},
	author       = {Sermpezis, Pavlos and Kotronis, Vasileios and Dainotti, Alberto and Dimitropoulos, Xenofontas},
	year         = 2018,
	journal      = {ACM SIGCOMM Computer Communication Review},
	publisher    = {ACM New York, NY, USA},
	volume       = 48,
	number       = 1,
	pages        = {64--69}
}

@misc{pooled-staking,
	title        = {Pooled staking},
	author       = {{Ethereum Foundation}},
	url          = {https://ethereum.org/en/staking/pools/},
	note         = {Accessed 21 April 2025.}
}

@article{heimbach2024deanonymizing,
	title        = {Deanonymizing ethereum validators: The p2p network has a privacy issue},
	author       = {Heimbach, Lioba and Vonlanthen, Yann and Villacis, Juan and Kiffer, Lucianna and Wattenhofer, Roger},
	year         = 2024,
	journal      = {arXiv preprint arXiv:2409.04366}
}

@misc{qrator2023q2,
	title        = {Q2 2023 DDoS attacks statistics and overview},
	author       = {Qrator Labs},
	year         = 2023,
	note         = {Accessed: 2025-06-03},
	howpublished = {\url{https://blog.qrator.net/en/q2-2023-ddos-attacks-statistics-and-overview_177/}}
}

@misc{qrator2023q1,
	title        = {Q1 2023 DDoS Attacks and BGP Incidents},
	author       = {Qrator Labs},
	year         = 2023,
	note         = {Accessed: 2025-06-03},
	howpublished = {\url{https://blog.qrator.net/en/q1-2023-ddos-attacks-and-bgp-incidents_171/}}
}

@misc{qrator2025q1,
	title        = {Q1 2025 DDoS, Bots and BGP Incidents Statistics and Overview},
	author       = {Qrator Labs},
	year         = 2025,
	note         = {Accessed: 2025-06-03},
	howpublished = {\url{https://blog.qrator.net/en/q1-2025-ddos-bots-and-bgp-incidents-statistics-and_211/}}
}

@misc{manrs2022bgp,
	title        = {BGP Security in 2021},
	author       = {{Mutually Agreed Norms for Routing Security (MANRS)}},
	year         = 2022,
	note         = {Accessed: 2025-06-03},
	howpublished = {\url{https://manrs.org/2022/02/bgp-security-in-2021/}}
}

@misc{cloudflare2024incident,
	title        = {Cloudflare 1.1.1.1 Incident on June 27, 2024},
	author       = {Bryton Herdes and Mingwei Zhang and Tanner Ryan},
	year         = 2024,
	note         = {Accessed: 2025-06-03},
	howpublished = {\url{https://blog.cloudflare.com/cloudflare-1111-incident-on-june-27-2024/}}
}

@misc{cloudflare2021facebook,
	title        = {Understanding how Facebook disappeared from the Internet},
	author       = {Celso Martinho and Tom Strickx},
	year         = 2021,
	note         = {Accessed: 2025-06-03},
	howpublished = {\url{https://blog.cloudflare.com/october-2021-facebook-outage/}}
}

@article{milolidakis2023effectiveness,
	title        = {On the effectiveness of bgp hijackers that evade public route collectors},
	author       = {Milolidakis, Alexandros and B{\"u}hler, Tobias and Wang, Kunyu and Chiesa, Marco and Vanbever, Laurent and Vissicchio, Stefano},
	year         = 2023,
	journal      = {IEEE Access},
	publisher    = {IEEE},
	volume       = 11,
	pages        = {31092--31124}
}

@misc{orange2024bgp,
	title        = {Hacker hijacks Orange Spain RIPE account to cause BGP havoc},
	author       = {Lawrence Abrams},
	year         = 2024,
	note         = {Accessed: 2025-06-03},
	howpublished = {\url{https://www.bleepingcomputer.com/news/security/hacker-hijacks-orange-spain-ripe-account-to-cause-bgp-havoc/}}
}

@article{10.1145/3224424,
	title        = {Dandelion++: Lightweight Cryptocurrency Networking with Formal Anonymity Guarantees},
	author       = {Fanti, Giulia and Venkatakrishnan, Shaileshh Bojja and Bakshi, Surya and Denby, Bradley and Bhargava, Shruti and Miller, Andrew and Viswanath, Pramod},
	year         = 2018,
	month        = jun,
	journal      = {Proc. ACM Meas. Anal. Comput. Syst.},
	publisher    = {Association for Computing Machinery},
	address      = {New York, NY, USA},
	volume       = 2,
	number       = 2,
	doi          = {10.1145/3224424},
	url          = {https://doi.org/10.1145/3224424},
	issue_date   = {June 2018},
	articleno    = 29,
	numpages     = 35
}

@inproceedings{mahajan2002understanding,
	title        = {Understanding BGP misconfiguration},
	author       = {Mahajan, Ratul and Wetherall, David and Anderson, Thomas},
	year         = 2002,
	booktitle    = {Proceedings of the 2002 conference on Applications, technologies, architectures, and protocols for computer communications},
	pages        = {3--16},
	organization = {ACM}
}

@misc{mevboost.pics,
	title        = {Mevboost.pics - Open Data},
	author       = {Wahrstätter, Toni},
	year         = 2023,
	url          = {https://mevboost.pics/data.html},
	note         = {Accessed April 12 2025.}
}

@misc{pyasn,
	title        = {pyasn: Python IP address to Autonomous System Number lookup module},
	author       = {Asghari, Hadi and Noroozian, Arman},
	year         = 2014,
	url          = {https://github.com/hadiasghari/pyasn},
	note         = {Accessed August 25 2024.}
}

@misc{ipinfo-lite,
	title        = {{IPInfo Lite}},
	author       = {{IPInfo}},
	year         = 2026,
	url          = {https://ipinfo.io/lite},
	note         = {Accessed January 4 2026.}
}

@article{neuder2021low,
	title        = {Low-cost attacks on Ethereum 2.0 by sub-1/3 stakeholders},
	author       = {Neuder, Michael and Moroz, Daniel J and Rao, Rithvik and Parkes, David C},
	year         = 2021,
	journal      = {arXiv preprint arXiv:2102.02247}
}

@inproceedings{neu2021ebb,
	title        = {Ebb-and-flow protocols: A resolution of the availability-finality dilemma},
	author       = {Neu, Joachim and Tas, Ertem Nusret and Tse, David},
	year         = 2021,
	booktitle    = {2021 IEEE Symposium on Security and Privacy (SP)},
	pages        = {446--465},
	organization = {IEEE}
}

@inproceedings{three-attacks-on-pos,
	title        = {Three Attacks on Proof-of-Stake Ethereum},
	author       = {Schwarz-Schilling, Caspar and Neu, Joachim and Monnot, Barnab{\'e} and Asgaonkar, Aditya and Tas, Ertem Nusret and Tse, David},
	year         = 2022,
	booktitle    = {Financial Cryptography and Data Security},
	publisher    = {Springer International Publishing},
	address      = {Cham},
	isbn         = {978-3-031-18283-9},
	editor       = {Eyal, Ittay and Garay, Juan}
}

@article{miller2015discovering,
	title        = {Discovering bitcoin’s public topology and influential nodes.(2015)},
	author       = {Miller, Andrew and Litton, James and Pachulski, Andrew and Gupta, Neal and Levin, Dave and Spring, Neil and Bhattacharjee, Bobby and others},
	year         = 2015,
	pages        = 54
}

@inproceedings{two-more-attacks,
	title        = {Two More Attacks on Proof-of-Stake GHOST/Ethereum},
	author       = {Neu, Joachim and Tas, Ertem Nusret and Tse, David},
	year         = 2022,
	booktitle    = {Proceedings of the 2022 ACM Workshop on Developments in Consensus},
	location     = {Los Angeles, CA, USA},
	publisher    = {Association for Computing Machinery},
	address      = {New York, NY, USA},
	series       = {ConsensusDay '22},
	pages        = {43–52},
	doi          = {10.1145/3560829.3563560},
	isbn         = 9781450398794,
	url          = {https://doi.org/10.1145/3560829.3563560},
	numpages     = 10
}

@inproceedings{pos-under-scrutiny,
	title        = {Ethereum Proof-of-Stake under Scrutiny},
	author       = {Pavloff, Ulysse and Amoussou-Guenou, Yackolley and Tucci-Piergiovanni, Sara},
	year         = 2023,
	booktitle    = {Proceedings of the 38th ACM/SIGAPP Symposium on Applied Computing},
	location     = {Tallinn, Estonia},
	publisher    = {Association for Computing Machinery},
	address      = {New York, NY, USA},
	series       = {SAC '23},
	pages        = {212–221},
	doi          = {10.1145/3555776.3577655},
	isbn         = 9781450395175,
	url          = {https://doi.org/10.1145/3555776.3577655},
	numpages     = 10
}

@inproceedings{10646904,
	title        = {Byzantine Attacks Exploiting Penalties in Ethereum PoS},
	author       = {Pavloff, Ulysse and Amoussou-Guenou, Yackolley and Tucci-Piergiovanni, Sara},
	year         = 2024,
	booktitle    = {2024 54th Annual IEEE/IFIP International Conference on Dependable Systems and Networks (DSN)},
	volume       = {},
	number       = {},
	pages        = {53--65},
	doi          = {10.1109/DSN58291.2024.00020}
}

@article{10.1145/2659899,
	title        = {Why is it taking so long to secure internet routing?},
	author       = {Goldberg, Sharon},
	year         = 2014,
	month        = sep,
	journal      = {Commun. ACM},
	publisher    = {Association for Computing Machinery},
	address      = {New York, NY, USA},
	volume       = 57,
	number       = 10,
	pages        = {56–63},
	doi          = {10.1145/2659899},
	issn         = {0001-0782},
	url          = {https://doi.org/10.1145/2659899},
	issue_date   = {October 2014},
	numpages     = 8
}

@misc{cryptoeprint:2016/1010,
	title        = {Are We There Yet? On {RPKI}'s Deployment and Security},
	author       = {Yossi Gilad and Avichai Cohen and Amir Herzberg and Michael Schapira and Haya Shulman},
	year         = 2016,
	url          = {https://eprint.iacr.org/2016/1010},
	howpublished = {Cryptology {ePrint} Archive, Paper 2016/1010}
}

@techreport{rekhter2006border,
	title        = {A border gateway protocol 4 (BGP-4)},
	author       = {Rekhter, Yakov and Li, Tony and Hares, Susan},
	year         = 2006
}

@article{bgp-misconfiguration,
	title        = {Understanding BGP misconfiguration},
	author       = {Mahajan, Ratul and Wetherall, David and Anderson, Tom},
	year         = 2002,
	month        = aug,
	publisher    = {Association for Computing Machinery},
	address      = {New York, NY, USA},
	volume       = 32,
	number       = 4,
	pages        = {3–16},
	doi          = {10.1145/964725.633027},
	issn         = {0146-4833},
	url          = {https://doi.org/10.1145/964725.633027},
	issue_date   = {October 2002},
	numpages     = 14
}

@article{bailey2012menlo,
	title        = {The menlo report},
	author       = {Bailey, Michael and Dittrich, David and Kenneally, Erin and Maughan, Doug},
	year         = 2012,
	journal      = {IEEE Security \& Privacy},
	publisher    = {IEEE},
	volume       = 10,
	number       = 2,
	pages        = {71--75}
}

@inproceedings{boneh2020single,
	title        = {Single secret leader election},
	author       = {Boneh, Dan and Eskandarian, Saba and Hanzlik, Lucjan and Greco, Nicola},
	year         = 2020,
	booktitle    = {Proceedings of the 2nd ACM Conference on Advances in Financial Technologies},
	pages        = {12--24}
}

@inproceedings{catalano2023efficient,
	title        = {Efficient and universally composable single secret leader election from pairings},
	author       = {Catalano, Dario and Fiore, Dario and Giunta, Emanuele},
	year         = 2023,
	booktitle    = {IACR International Conference on Public-Key Cryptography},
	pages        = {471--499},
	organization = {Springer}
}

@article{whisk,
	title        = {Whisk: A practical shuffle-based SSLE protocol for Ethereum},
	author       = {Kadianakis, George and Drake, Justin and Feist, Dankrad and Herold, Gottfried and Khovratovich, Dmitry and Maller, Mary and Simkin, Mark},
	year         = 2022,
	journal      = {Ethereum Research},
	url          = {https://ethresear.ch/t/whisk-a-practical-shuffle-based-ssle-protocol-for-ethereum/11763},
	note         = {Online; accessed June 4, 2025}
}

@misc{own-asn,
	title        = {{How I set up my own Autonomous System}},
	author       = {Swer, Daryll},
	year         = 2022,
	note         = {Accessed: fo2025-06-05},
	howpublished = {\textit{daryllswer.com}}
}

@article{apnic-fees,
	title        = {How much does it cost?},
	author       = {APNIC},
	year         = 2025,
	journal      = {APNIC},
	url          = {https://www.apnic.net/get-ip/apnic-membership/how-much-does-it-cost/}
}

@article{nick-isp,
	title        = {Setting up a Personal ASN},
	author       = {Bouwhuis, Nick},
	year         = 2023,
	journal      = {nick.bouwhuis.net},
	url          = {https://nick.bouwhuis.net/posts/2023-02-12-setting-up-a-personal-asn/}
}

@article{lido-operators,
	title        = {Ethereum Node Operators},
	author       = {Lido},
	year         = 2025,
	journal      = {Lido Node Operator Portal},
	url          = {https://operatorportal.lido.fi/lido-operators-database/ethereum-node-operators},
	note         = {Online; accessed 5 June 2025}
}

@misc{wired2014bgphijack,
	title        = {Hacker Redirects Traffic From 19 Internet Providers to Steal Bitcoins},
	author       = {Andy Greenberg},
	year         = 2014,
	note         = {Accessed: 2025-06-06},
	howpublished = {\url{https://www.wired.com/2014/08/isp-bitcoin-theft/}}
}

@inproceedings{heo2023partitioning,
	title        = {Partitioning Ethereum without Eclipsing It.},
	author       = {Heo, Hwanjo and Woo, Seungwon and Yoon, Taeung and Kang, Min Suk and Shin, Seungwon},
	year         = 2023,
	booktitle    = {NDSS}
}

@inproceedings{hkzg_15,
	title        = {Eclipse Attacks on Bitcoin's Peer-to-Peer Network},
	author       = {Ethan Heilman and Alison Kendler and Aviv Zohar and Sharon Goldberg},
	year         = 2015,
	booktitle    = {24th {USENIX} Security Symposium, {USENIX} Security 15, Washington, D.C., USA, August 12-14, 2015.},
	pages        = {129--144},
	url          = {https://www.usenix.org/conference/usenixsecurity15/technical-sessions/presentation/heilman},
	bibsource    = {dblp computer science bibliography, http://dblp.org}
}

@misc{celer,
	title        = {Celer Bridge incident analysis},
	howpublished = {\url{https://www.coinbase.com/blog/celer-bridge-incident-analysis}},
	key          = {celer}
}

@inproceedings{tran2020stealthier,
	title        = {A stealthier partitioning attack against bitcoin peer-to-peer network},
	author       = {Tran, Muoi and Choi, Inho and Moon, Gi Jun and Vu, Anh V and Kang, Min Suk},
	year         = 2020,
	booktitle    = {2020 IEEE Symposium on Security and Privacy (SP)},
	pages        = {894--909},
	organization = {IEEE}
}

@inproceedings{saad2019partitioning,
	title        = {Partitioning attacks on bitcoin: Colliding space, time, and logic},
	author       = {Saad, Muhammad and Cook, Victor and Nguyen, Lan and Thai, My T and Mohaisen, Aziz},
	year         = 2019,
	booktitle    = {2019 IEEE 39th international conference on distributed computing systems (ICDCS)},
	pages        = {1175--1187},
	organization = {IEEE}
}

@inproceedings{sabre,
	title        = {{SABRE: Protecting Bitcoin against Routing Attacks}},
	author       = {M. Apostolaki and Marti, Gian and M\"uller, Jan and Vanbever, Laurent},
	year         = 2019,
	booktitle    = {Proceedings of the 26th Network and Distributed System Security Symposium (NDSS'19)}
}

@article{sun2021securing,
	title        = {Securing internet applications from routing attacks},
	author       = {Sun, Yixin and Apostolaki, Maria and Birge-Lee, Henry and Vanbever, Laurent and Rexford, Jennifer and Chiang, Mung and Mittal, Prateek},
	year         = 2021,
	journal      = {Communications of the ACM},
	publisher    = {ACM New York, NY, USA},
	volume       = 64,
	number       = 6,
	pages        = {86--96}
}

@misc{myetherwallet,
	title        = {A \$152,000 Cryptocurrency Theft Just Exploited A Huge 'Blind Spot' In Internet Security},
	howpublished = {https://www.forbes.com/sites/thomasbrewster/2018/04/24/a-160000-ether-theft-just-exploited-a-massive-blind-spot-in-internet-security/?sh=35607bc85e26},
	key          = {myetherwallet}
}

@article{vyzovitis2020gossipsub,
	title        = {GossipSub: Attack-resilient message propagation in the Filecoin and ETH2. 0 networks},
	author       = {Vyzovitis, Dimitris and Napora, Yusef and McCormick, Dirk and Dias, David and Psaras, Yiannis},
	year         = 2020,
	journal      = {arXiv preprint arXiv:2007.02754}
}
